\documentclass[a4paper,onecolumn,unpublished,11pt]{quantumarticle}
\pdfoutput=1
\usepackage{PTRSA_env}

\usepackage{xfrac}
\usepackage{hyperref}
\usepackage[page]{appendix}

\begin{document}

    \title{\color{quantumviolet}Corrected Bell and Noncontextuality Inequalities \newline for Realistic Experiments}

\author{Kim Vallée\texorpdfstring{$^*$}{*}}
\email{kim.vallee@lip6.fr}
\orcid{0000-0002-4743-3984}
\affiliation{Sorbonne Université, CNRS, LIP6, F-75005 Paris, France}
\author{Pierre-Emmanuel Emeriau\texorpdfstring{$^*$}{*}}
\email{pe.emeriau@quandela.com}
\orcid{0000-0001-5155-1783}
\affiliation{Quandela, 7 Rue Léonard de Vinci, 91300 Massy, France}
\author{Boris Bourdoncle}
\affiliation{Quandela, 7 Rue Léonard de Vinci, 91300 Massy, France}
\orcid{0000-0002-2686-833X}
\author{Adel Sohbi}
\affiliation{ORCA Computing, Toronto, M6P 3T1, Canada}
\orcid{0000-0002-1275-9722}
\author{Shane Mansfield}
\affiliation{Quandela, 7 Rue Léonard de Vinci, 91300 Massy, France}
\orcid{0000-0002-2231-2422}
\author{Damian Markham}
\affiliation{Sorbonne Université, CNRS, LIP6, F-75005 Paris, France}
\orcid{0000-0003-3111-7976}

\maketitle

\begin{abstract}
Contextuality is a feature of quantum correlations.
It is crucial from a foundational perspective as a nonclassical phenomenon, and from an applied perspective as a resource for quantum advantage.
It is commonly defined in terms of hidden variables, for which it forces a contradiction with the assumptions of parameter-independence and determinism.
The former can be justified by the empirical property of non-signalling or non-disturbance, and the latter by the empirical property of measurement sharpness.
However, in realistic experiments neither empirical property holds exactly,
which leads to possible objections to contextuality as a form of nonclassicality,
and potential vulnerabilities for supposed quantum advantages.
We introduce measures to quantify both properties, and introduce quantified relaxations of the corresponding assumptions.
We prove the continuity of a known measure of contextuality, the contextual fraction, which ensures its robustness to noise.
We then bound the extent to which these relaxations can account for contextuality,
via corrections terms to the contextual fraction (or to any noncontextuality inequality),
culminating in a notion of genuine contextuality, which is robust to experimental imperfections. We then show that our result is general enough to apply or relate to a variety of established results and experimental setups. 
\end{abstract}

\section{Introduction}

Contextuality is well-studied as a nonclassical feature of the empirical predictions of quantum mechanics,
which may be considered as a generalisation of Bell nonlocality \cite{belleinstein1964}.
First studied by Bell \cite{bell1966} and Kochen and Specker \cite{KochenSpecker1967},
there now exist a number of general, structural frameworks for treating contextuality,
including the sheaf \cite{abramsky2011sheaf}, graph \cite{csw},
hypergraph \cite{afls}, and contextuality-by-default \cite{dzhafarov2016context} frameworks.
The study of contextuality has a strong foundational interest, but also led to a range of recent results that establish links between contextuality and quantum advantage in the aforementioned frameworks
\cite{raussendorf2013contextuality,howard2014contextuality,abramsky2017contextual,bermejo2017contextuality,abramsky2017quantum,mansfield2018quantum,karanjai2018contextuality,bowles_contextuality_2023,fyrillas_certified_2023}\footnotemark,
which makes it an important phenomenon for applications in quantum information and computation.

\footnotetext{In particular, the randomness certification protocol demonstrated in \cite{fyrillas_certified_2023}, which is robust to cross-talk effects inherent to the experiment, makes use of techniques introduced in the present work.}

As it is usually presented, the physical consequences of contextuality are such that for any hidden variable theory reproducing the observed behaviour it forces one to give up on either \textit{determinism} or \textit{parameter independence}.
In this way, very roughly speaking, we could say that contextuality prevents the interpretation of measurements as revealing pre-determined values (determinism)---at least in a way that does not disturb the supposed underlying state of the system (parameter independence).
The structural formalisms mentioned above admit equivalent definitions, but which shift the focus away from hidden variables, and bring into play new techniques and methodologies for analysing contextuality. However, in attempting to grasp the physical meaning of contextuality the hidden variable description is nevertheless often the most insightful.
Through the article we will use a quantitative measure of contextuality, known as the Contextual Fraction ($\CF$) \cite{abramsky2011sheaf},
which has been shown to generalise Bell and noncontextual inequalities \cite{abramsky2017contextual}. \footnote{The $\CF$ can be viewed as a generalisation of the non-local content defined in \cite{Barrett2006LF}. A related notion of noncontextual content \cite{amselem2012Experimental}, which appeared later, corresponds to $1 - \CF$; in other words it is the noncontextual fraction.}

However, in realistic, noisy experiments, the validity of the assumptions of \textit{determinism} and \textit{parameter independence} can be called into question.
As we will discuss, and as has been pointed out elsewhere (e.g.\ \cite{spekkens2014status}), beyond certain noise thresholds a contextual experiment can admit explanations in line with classical intuitions. 
While noise does not directly feature in hidden variable models, a standard justification for hidden variable assumptions is that they reflect some empirical property \cite{leifer2014quantum}.
So if noise affects those properties it will also weaken the validity of the assumptions.

In the case of \textit{determinism}, the analogous empirical property is \textit{sharpness} (usually of measurements).
In theory at least, classical and quantum physics allow for measurements that are perfectly sharp:
e.g.\ if the measurement is immediately repeated it returns the same outcome \cite{chiribella2014measurement}.
An extreme example of a dichotomic unsharp measurement would be one which independent of the state returns outcomes $0$ or $1$ with equal probability.

While this is perfectly reasonable as a measurement,
it would be unreasonable to assume that such a measurement is capable of revealing pre-determined values of some properties of the state it is applied to.
In practice, realistic measurements may be sharp to a good approximation but will never be {perfectly} sharp.

In the case of \textit{parameter independence} the analogous empirical property is \textit{no-signalling} or \textit{no-disturbance}.
In Bell-type scenarios, no-signalling is the property that any party's measurement choice does not influence other parties' measurement outcomes.
In contextuality scenarios like those of \cite{abramsky2011sheaf,csw,afls,dzhafarov2016context}, we do not need to assume different spacelike separated parties but may still speak of a generalised no-signalling property sometimes referred to as \textit{no-disturbance}.
In theory at least, the empirical behaviours of systems that obey classical or quantum physics should be perfectly non-signalling
in this generalised sense.
In practice, however, no realistic empirical model can ever be \textit{perfectly} non-signalling.
This could simply be due to finite statistics, or have to do with systematic errors such as imperfect shielding or experimental crosstalk.
Our work is concerned with systematic errors and the properties of the behaviour underlying given empirical data.
For methods to deal with the problem of finite statistics in experimental trials, we refer the reader to \cite{zhang2011Asymptotically,zhang2013Efficient} when testing local realism, and to \cite{dupuis2020Entropy,lin2018Deviceindependent} for device-independent information processing.

\begin{table}[!ht]
\centering
    \begin{tabular}{ l|l } 
        \multicolumn{1}{c|}{Empirical property} & \multicolumn{1}{c}{Hidden-variable property} \\
    \hline \hline
        No-Signalling (NS) & Parameter Independence (PI) \\
        Sharpness & Outcome Determinism (OD) 
    \end{tabular}
    \label{tab:empHV}
    \caption{Comparison between characteristics of the empirical data and its counterpart at the hidden-variable level.}
\end{table}

So if we never observe perfectly sharp nor non-signalling empirical data, why should we expect an underlying hidden variable or physical description of the system be perfectly deterministic and parameter independent?
We propose to quantify unsharpness and signalling, and to admit hidden variable models that are nondeterministic and parameter dependent to a corresponding extent.
Contextuality should only be considered robust to the above objections if it cannot be accounted for by this additional flexibility afforded to hidden variable explanations.
In other words, any attempt to explain robust contextuality would necessitate \emph{more} non-determinism or \emph{more} disturbance than is empirically justified.

\paragraph{Relationship to Contextuality-by-Default (CbD)}
The CbD approach to contextuality \cite{dzhafarov2016context} is a general (hidden-variable) model-independent treatment of contextuality in the presence of signalling,
in part motivated by applications to empirical data arising both in physical \cite{kujala2015necessary} and non-physical settings, e.g.\ in linguistics \cite{wang2021analysing,wang2021quantum} or psychology \cite{cervantes2018snow,basieva2019true}.
A key component of CbD analysis is the notion of a maximal coupling of empirical data.
Although it is not the usual perspective of CbD, maximal couplings can be viewed as a kind of hidden variable model in which the amount of parameter dependence is constrained to match the amount of signalling present in the empirical data, wherever measurements appear in multiple contexts.
At a conceptual level there is thus a clear similarity between CbD and the present approach.
There are also some clear distinctions.
We wish to reason about which physical mechanisms could underlie a given empirical behaviour.
For that reason, from the outset, we explicitly anchor the foundations of our approach in terms of hidden variable models, as they provide a broad framework for hypothesising and reasoning about such mechanisms.
By design, it is therefore clearer how to extract a physical meaning from an observation of genuine contextuality with our approach.\footnotemark 
In contrast to CbD, the present approach admits the degree of parameter dependence to be bounded by empirical signalling in a flexible manner -- in particular without requiring the bounds to be saturated, allowing for different kinds of hidden variable explanation.
In any case the quantification of signalling and parameter dependence in both approaches are different.
Our quantification methods are motivated by natural connection or analogy to the contextual fraction \cite{abramsky2017contextual}.
This is known to coincide \cite{cervantes2023note,camillo2023measures} in some specific scenarios with measures used within the CbD approach \cite{kujala2019measures},
which however are only defined for dichotomic measurements.
Finally, while the CbD approach applies to signalling, we will also bring into play considerations of sharpness.

\footnotetext{The approach may have significance for reasoning about mechanisms underlying empirical data in other settings too, but here our primary focus is on their well-established application to physical phenomena.}

\paragraph{Relationship to other previous works}
Previous analyses have focused specifically on robustness to unsharpness \cite{kunjwal2015kochen,kunjwal2019beyond,kunjwal2018statistical}, or to signalling \cite{winter2014does,dzhafarov2016context}.
Others have considered generalised notions of hidden variable models that allow extra expressivity through state updating \cite{kirchmair2009state,kleinmann2011memory}.
The work of \cite{hall2011Relaxed} takes into account both signalling and unsharpness, although the analysis is limited to the Bell-CHSH scenario; in particular our approach retrieves those results and extends them to any scenario. Finally, one work had similar motivations on closeness of behaviours \cite{horodecki2015Axiomatica}, however their approach does not take into account hidden variables, and focus on the resources and other measures of contextuality. The continuity of various measures of contextuality have been investigated in \cite{amaral2017geometrical} and other asymptotic limits in \cite{horodecki2015Axiomatica} but the contextual fraction has never been proven to be continuous before this work.

Our aim is to provide a simplified, holistic approach that is robust to both objections.
We show that experiment-friendly
\cite{fyrillas_certified_2023} inequalities that straightforwardly extend the contextual fraction measure \cite{abramsky2017contextual} and test for witnessing \textit{genuine} contextuality can be obtained by setting lower bounds on the admissible fraction of nondeterminism $\eta$ and parameter-dependence $\sigma$.

\section{General framework for contextuality}

Here we briefly summarise some of the main ideas from \cite{abramsky2011sheaf,abramsky2017contextual}. An extended introduction can be found in the Supplemental Material.

\subsection{Ingredients à la Abramsky-Brandenburger}

\hspace{10pt}\textit{Measurement scenario.}--- An abstract description of an experimental setup is described by a measurement scenario $\XMO$ where 
\begin{enumerate*}[label=(\roman*)]
    \item $X$ is a finite set of measurement labels;
    \item $\Mc \subseteq \Pc(X)$ is the set of all maximal contexts, \ie sets of compatible measurements, and forms a cover of $X$;
    \item $O = (O_x)_{x \in X}$ where $O_x$ is the outcome set corresponding to measurement $x \in X$.
\end{enumerate*}

\textit{Empirical behaviours.}---
Given the description of the experimental setup, either calculating theoretical predictions for admissible joint outcomes or performing repeated runs of the experiment with varying choices of measurement context and recording the frequencies of the corresponding joint events results in a probability table which is formalised as an empirical model. Formally, an empirical model, or behaviour, $e$ on $\XMO$ is a family $e = (e_C)_{C \in \Mc}$ where $e_C$ is a probability distribution on the corresponding joint outcome space. 

\textit{Non-signalling.}--- 
A behaviour $e$ is non-signalling when for any two contexts $C_1$ and $C_2$, $e_{C_1}|_{C_1 \cap C_2} = e_{C_2}|_{C_1 \cap C_2}$. The notation ${e_C}|_{U}$ with stands for the marginalisation of the probability distribution ${e_C}$ to $U\subseteq C$: for $t \in O_U$, ${e_C}|_{U}(t) := \sum_{s\in O_C,{s}|_{U} =t }  e_C (s)$, where $s|_U$ is simply the function restriction of $s$ to the domain $U$ and $O_U := \prod_{x\in U} O_x$. 

\textit{(Non)contextuality.}--- 
Informally, an empirical behaviour $e$ is said to be noncontextual whenever context-wise predictions can all be obtained as the marginals of one probability distribution over global value assignments. 

Note at this point that the statement concerns only the empirical model, and we have not referred to any underlying hidden variable model.
The relationship with hidden variable models was made explicit in \cite{abramsky2011sheaf} for the case of perfectly parameter-independent hidden variable models (see Proposition \ref{prop:sheafequiv}).
Thus the Abramsky--Brandenburger approach is equivalent to, yet neatly abstracts away from, hidden-variable models. It also leads to convenient tools like the contextual fraction to quantify contextuality \cite{abramsky2017contextual}, which we recap next.
Establishing an analogous relationship for noisy data, and a convenient method for quantifying contextuality in a noise-robust manner contextual is the main contibution of this work.

\subsection{Quantifying contextuality and signalling}

By definition $O_X$ is the set of global assignments of an outcome to each measurement. Let $n \defeq |O_X|$ be the number of global assignments and $m \defeq \sum_{C \in \Mc} |O_C| $ the number of total `local' (or context-wise) assignments ranging over all contexts. We use bold notation for vectors. Local assignments can be listed as: $\left\{ \langle C,\bm s \rangle \text{ s.t. } C \in \Mc \text{ and } \bm s \in O_C \right\}$. Then the incidence matrix $\Mrm$ which records the possible restrictions from global assignments $g$ to local assignments $\langle C,\bm s \rangle$ is a $m \times n$ (0,1)-matrix defined as:
\begin{equation}
    \Mrm[\langle C, \bm s \rangle,g] \defeq 
    \begin{cases}
    1 \text{ if } \bm g|_C = \bm s \\
    0 \text{ otherwise} \Mdot
    \end{cases}
    \label{eq:incidencemat}
\end{equation}
To understand its action, consider the columns of $\Mrm$ as listing the global assignments $\bm g \in O_X$. In particular, notice that rows are labelled by local assignments $\bm s$, and that for any given row and column, $\Mrm$ assigns the value $1$ whenever the corresponding local and global assignments agree, $\bm g|_C = \bm s$ (and $0$ otherwise).

The empirical model $e$ can be represented as a vector $ \bm \vrm^e  \in \left[0,1\right]^m$ where for a given context $C \in \Mc$ and a local assignment $\bm s \in \Oc_C$, $\bm \vrm^e[\langle C, \, \bm s \rangle] = e_C(\bm s)$. This is the flattened version of the tables that are usually used to represent empirical models.

\begin{figure}[ht]
\centering
    \scalebox{1}{\makeatletter
\tikzoption{canvas is plane}[]{\@setOxy#1}
\def\@setOxy O(#1,#2,#3)x(#4,#5,#6)y(#7,#8,#9)%
  {\def\tikz@plane@origin{\pgfpointxyz{#1}{#2}{#3}}%
   \def\tikz@plane@x{\pgfpointxyz{#4}{#5}{#6}}%
   \def\tikz@plane@y{\pgfpointxyz{#7}{#8}{#9}}%
   \tikz@canvas@is@plane
  }
\makeatother  

\begin{tikzpicture}[scale=1.]

\tikzstyle{vertex}=[circle, draw, inner sep=0pt, minimum size=3pt]
\newcommand{\vertex}{\node[vertex]}


\vertex[yellow!80,fill=yellow!80] (s1) at (2,-2.5,-5.25) {};
\vertex[yellow!80,fill=yellow!80] (s2) at (2,2.5,-5.25) {};
\vertex[yellow!80,fill=yellow!80] (s3) at (5.25,-2.5,-2) {};
\vertex[yellow!80,fill=yellow!80] (s4) at (5.25,2.5,-2) {};
\vertex[yellow!80,fill=yellow!80] (s5) at (-1.25,2.5,-2) {};
\vertex[yellow!80,fill=yellow!80] (s6) at (-1.25,-2.5,-2) {}; %
\vertex[yellow!80,fill=yellow!80] (s7) at (2,2.5,1.25) {}; 
\vertex[yellow!80,fill=yellow!80, text=red] (s8) at (2,-2.5,1.25) {}; %

\draw[draw=none,fill=yellow!10,opacity=0.3] (2,-2.5,-5.25) -- (2,2.5,-5.25) -- (0,0,-4) -- (2,-2.5,-5.25);
\draw[draw=none,fill=yellow!10,opacity=0.3] (2,-2.5,-5.25) -- (2,2.5,-5.25) -- (4,0,-4) -- (2,-2.5,-5.25);

\draw[draw=none,fill=yellow!10,opacity=0.3] (2,2.5,-5.25) -- (-1.25,2.5,-2) -- (0,0,-4) -- (2,2.5,-5.25);
\draw[draw=none,fill=yellow!10,opacity=0.3] (2,2.5,-5.25) -- (5.25,2.5,-2) -- (4,0,-4) -- (2,2.5,-5.25);

\draw[draw=none,fill=yellow!10,opacity=0.3] (0,0,-4) -- (-1.25,-2.5,-2) -- (2,-2.5,-5.25) -- (0,0,-4);
\draw[draw=none,fill=yellow!10,opacity=0.3] (4,0,-4) -- (5.25,-2.5,-2) -- (2,-2.5,-5.25) -- (4,0,-4);

\draw[draw=none,fill=yellow!10,opacity=0.3] (-1.25,-2.5,-2) -- (-1.25,2.5,-2) -- (0,0,-4) -- (-1.25,-2.5,-2);

\draw[draw=none,fill=yellow!10,opacity=0.3] (5.25,-2.5,-2) -- (5.25,2.5,-2) -- (4,0,-4) -- (5.25,-2.5,-2);

\draw[dashed, opacity=.3]  (2,-2.5,-5.25) -- (2,2.5,-5.25) -- (0,0,-4) -- (2,-2.5,-5.25);
\draw[dashed, opacity=.3]  (2,-2.5,-5.25) -- (4,0,-4) -- (2,2.5,-5.25);

\draw[opacity=.3]  (2,2.5,-5.25) -- (-1.25,2.5,-2);
\draw[opacity=.3]  (2,2.5,-5.25) -- (5.25,2.5,-2);
\draw[dashed, opacity=.3] (0,0,-4) -- (-1.25,2.5,-2);

\draw[dashed, opacity=.3]  (0,0,-4) -- (-1.25,-2.5,-2);
\draw[dashed, opacity=.3]  (-1.25,-2.5,-2) -- (2,-2.5,-5.25);

\draw[dashed, opacity=.3]  (4,0,-4) -- (5.25,-2.5,-2) -- (2,-2.5,-5.25);

\draw[dashed, opacity=.3]  (5.25,2.5,-2) -- (4,0,-4);

\begin{scope}[canvas is plane={O(0,0,0)x(1,0,0)y(0,0,-1)},transform shape]
\draw[quantumturquoise!20,fill=quantumturquoise!20] (0,4) -- (2,5.25) -- (4,4) -- (5.25,2) -- (4,0) -- (2,-1.25) -- (0,0) -- (-1.25,2) -- (0,4);

\draw[quantumturquoise!40,fill=quantumturquoise!40]  (0,0) rectangle (4,4);

\vertex[quantumturquoise!100,fill=quantumturquoise!100] (p1) at (0,0) {};
\vertex[quantumturquoise!100,fill=quantumturquoise!100] (p2) at (0,4) {};
\vertex[quantumturquoise!100,fill=quantumturquoise!100] (p3) at (4,0) {};
\vertex[quantumturquoise!100,fill=quantumturquoise!100] (p4) at (4,4) {};

\vertex[quantumturquoise!60,fill=quantumturquoise!60] (p5) at (2,5.25) {};
\vertex[quantumturquoise!60,fill=quantumturquoise!60] (p6) at (5.25,2) {};
\vertex[quantumturquoise!60,fill=quantumturquoise!60] (p7) at (2,-1.25) {};
\vertex[quantumturquoise!60,fill=quantumturquoise!60] (p8) at (-1.25,2) {};

\node (L) at (2,2) {\Large $\mathcal NC$};
\node (NS) at (-.5,2) {\Large $\mathcal NS$};

\node[above left,rotate=-10] (eNS) at (1.5,4.5) {\Large \color{quantumdarkgray} $e^\text{NS}$};
\draw[dashed, quantumgray!80, thick] (2,5.25) -- (1.16,4);
\filldraw[quantumdarkgray!80] (1.5,4.5) circle (2pt);
\draw[<->,quantumviolet,thick] (1.8,4.36) -- (1.46,3.85);
\node[rotate=-30] (NCF) at (2.7,3.25) {\Large \color{quantumviolet}$\NCF(e^\text{NS})$};
\node[thick,draw=quantumdarkgray!80,cross out,inner sep=0pt,minimum width=2.5pt,minimum height=2.5pt] (eNC) at (1.16,4) {};
\node[thick,draw=quantumdarkgray!80,cross out,inner sep=0pt,minimum width=2.5pt,minimum height=2.5pt] (eNC) at (2,5.25) {};
\end{scope}

\draw[dashed, quantumdarkrose, thick] (2,2.5,-5.25) -- (1.5,0,-4.5);
\node[thick,draw=quantumdarkrose,cross out,inner sep=0pt,minimum width=1pt,minimum height=1pt] (eNC) at (1.5,0,-4.5) {};
\node[thick,draw=quantumdarkrose,cross out,inner sep=0pt,minimum width=1pt,minimum height=1pt] (eNC) at (2,2.5,-5.25) {};
\filldraw[quantumdarkrose] (1.735,1,-4.5) circle (2pt);
\node[left] (e) at (1.735,1,-4.5) { \color{quantumdarkrose} $e$};
\draw[<->,quantumblue,thick] (2.02,1,-4.5) -- (1.8,0,-4.36);
\node[right,rotate=-15] (NSF) at (1.8,0.1,-4.5) { \color{quantumblue}$\NSF(e)$};

\draw[draw=none,fill=yellow!20,opacity=0.3] (-1.25,-2.5,-2) -- (2,-2.5,1.25) -- (5.25,-2.5,-2) -- (2,-2.5,-5.25) -- (-1.25,-2.5,-2);

\draw[draw=none,fill=yellow!20,opacity=0.3] (-1.25,-2.5,-2) -- (-1.25,2.5,-2) -- (0,0,0) -- (-1.25,-2.5,-2);

\draw[draw=none,fill=yellow!20,opacity=0.3] (5.25,-2.5,-2) -- (5.25,2.5,-2) -- (4,0,0) -- (5.25,-2.5,-2);

\draw[draw=none,fill=yellow!20,opacity=0.3] (-1.25,2.5,-2) -- (0,0,0) -- (2,2.5,1.25) -- (-1.25,2.5,-2);
\draw[draw=none,fill=yellow!20,opacity=0.3] (-1.25,-2.5,-2) -- (0,0,0) -- (2,-2.5,1.25) -- (-1.25,-2.5,-2);

\draw[draw=none,fill=yellow!20,opacity=0.3] (5.25,2.5,-2) -- (4,0,0) -- (2,2.5,1.25) -- (5.25,2.5,-2);
\draw[draw=none,fill=yellow!20,opacity=0.3] (5.25,-2.5,-2) -- (4,0,0) -- (2,-2.5,1.25) -- (5.25,-2.5,-2);

\draw[draw=none,fill=yellow!20,opacity=0.3] (2,-2.5,1.25) -- (2,2.5,1.25) -- (0,0,0) -- (2,-2.5,1.25);
\draw[draw=none,fill=yellow!20,opacity=0.3] (2,-2.5,1.25) -- (2,2.5,1.25) -- (4,0,0) -- (2,-2.5,1.25);

\draw[draw=none,fill=yellow!20,opacity=0.3] (-1.25,2.5,-2) -- (2,2.5,1.25) -- (5.25,2.5,-2) -- (2,2.5,-5.25) -- (-1.25,2.5,-2);

\draw[opacity=0.3] (-1.25,-2.5,-2) -- (-1.25,2.5,-2) -- (0,0,0) -- (-1.25,-2.5,-2);

\draw[opacity=0.3] (5.25,-2.5,-2) -- (5.25,2.5,-2) -- (4,0,0) -- (5.25,-2.5,-2);

\draw[opacity=0.3] (0,0,0) -- (2,2.5,1.25) -- (-1.25,2.5,-2);
\draw[opacity=0.3] (0,0,0) -- (2,-2.5,1.25) -- (-1.25,-2.5,-2);

\draw[opacity=0.3] (4,0,0) -- (2,2.5,1.25) -- (5.25,2.5,-2);
\draw[opacity=0.3] (4,0,0) -- (2,-2.5,1.25) -- (5.25,-2.5,-2);

\draw[opacity=0.3] (2,-2.5,1.25) -- (2,2.5,1.25);

\node (MS1) at (1.75,-2.5,1.25) {\tiny $M_1$};
\node (MS2) at (1.75,2.5,1.25) {\tiny $M_2$};
\node (PR) at (1.75,0,1.25) {\tiny PR};
\node (D) at (-0.2,0,0) {\tiny D};

\end{tikzpicture}}
    \caption{A schematic representation of a cut of the signalling correlation polytope for the (2,2,2) Bell scenario with the no-signalling (NS) polytope and the noncontextual (NC) polytope inscribed in it. Vertices of the polytopes are highlighted: dark blue vertices correspond to NC sharp behaviours; light blue vertices to maximally contextual yet no-signalling behaviours (forming the NS polytope); and yellow vertices correspond to the signalling sharp behaviours. Note that contextual vertices of the non-signalling polytope, unlike noncontextual ones, are not vertices of the signalling polytope.
    The quantum set (not represented) in general corresponds to a convex set between the NS and NC polytopes.
    For a signalling behaviour $e$, its non-signalling fraction (maximal weight on the non-signalling part $e^\text{NS}$ in a convex decomposition) is represented. Likewise, the noncontextual fraction of $e^\text{NS}$ is represented. Labels displayed for some vertices are to be referred to in Figure~\ref{fig:PR_example}.}
    \label{fig:Spolytope}
\end{figure}
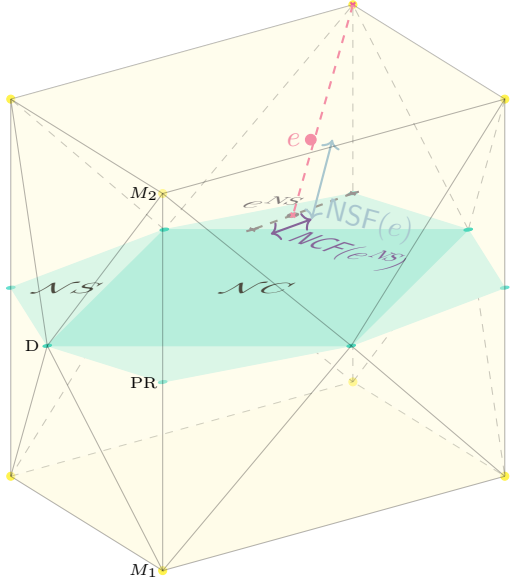

\textit{Contextual fraction.}---
Given empirical behaviours $e_1$ and $e_2$ on $\XMO$, $e = \lambda e_1 + (1-\lambda) e_2$ for $\lambda \in \left[0,1\right]$ is another valid empirical model. Instead of asking whether $e$ is contextual, we can wonder what fraction of $e$ can be explained noncontextually. This amounts to looking for a convex decomposition of the form $e = \lambda e^\text{NC} + (1-\lambda) e'$ with weight $\lambda$ on the noncontextual part. Maximising $\lambda$ yields the irreducible part of contextuality called the contextual fraction (see Figure~\ref{fig:Spolytope}): 
\begin{equation}
\label{eq:cf}
\begin{split}
    \CF(e) &\defeq 1-\NCF(e)\\
    &\defeq  1 - \max\{\lambda \mid \exists e^\text{NC} \text{ non-contextual for which } e' = \lambda e^\text{NC} + (1-\lambda) e'\} \, .
\end{split}
\end{equation}
It has a number of desirable properties \cite{abramsky2017contextual} including that it can be computed as a linear program: 
\leqnomode
\begin{flalign*}
    \label{prog:LP-CF}
    \tag*{(P-NCF)}
    \hspace{4cm}\left\{
    \begin{aligned}
        & \quad \text{Find } \bm \brm \in \R^n \\
        & \quad \text{maximising } \bm 1.\bm \brm \\
        & \quad \text{subject to:}  \\
        & \hspace{1cm} \begin{aligned}
            & \Mrm \bm \brm \leq \bm \vrm^e \\
            & \bm \brm \geq \bm 0 \Mdot
        \end{aligned}
    \end{aligned}
    \right. &&
\end{flalign*}
\reqnomode
This program computes the noncontextual fraction $\NCF(e)$ of $e$. We refer the interested readers to the Appendices for more details.

\textit{Signalling fraction.}---
In the same spirit, we introduce the signalling fraction $\SF(e)$ of a signalling behaviour $e$ by looking for a convex decomposition within the signalling polytope (see Figure~\ref{fig:Spolytope}) that maximises the weight of the non-signalling part:
\begin{equation*}
\begin{split}
    \SF(e) &\defeq 1-\NSF(e)\\
    &\defeq  1 - \max\{\lambda \mid \exists e^\text{NS} \text{ non-signalling for which } e' = \lambda e^\text{NS} + (1-\lambda) e'\} \, .
\end{split}    
\end{equation*}

This maximised weight can also take values in the interval $[0,1]$ and is called the signalling fraction\footnote{The signalling fraction originates from unpublished work of Samson Abramsky, Rui Soares Barbosa and SM.}. Just like the contextual fraction, the irreducible weight on the signalling part is the signalling fraction and can be computed with a linear program.
Idealised empirical behaviours predicted by quantum mechanics, in which contexts consist of perfectly commuting observables, will not be signalling.
However, in realistic settings noise can manifest itself to introduce some amount of signalling and
the signalling fraction defined above will quantify this as a property of the empirical data.

Just like the contextual fraction, the signalling fraction of any empirical behaviour can be efficiently computed through a linear program:
\leqnomode
\begin{flalign*}
    \label{prog:LP-SF}
    \tag*{(P-NSF)}
    \hspace{4cm}\left\{
    \begin{aligned}
        & \quad \text{Find } \bm \brm \in \R^n \\
        & \quad \text{maximising } \bm 1.\bm \brm \\
        & \quad \text{subject to:}  \\
        & \hspace{1cm} \begin{aligned}
            & \Mrm \bm \brm \leq \bm \vrm^e \\
            & \Mrm \bm \brm \geq 0 \Mdot
        \end{aligned}
    \end{aligned}
    \right. &&
\end{flalign*}
\reqnomode
Again this program is computing the no-signalling fraction $\NSF(e) \defeq 1 - \SF(e)$ of $e$.
Further details can be found in the Appendices.

\subsection{Relaxing parameter independence and outcome determinism}

\hspace{10pt}\textit{Hidden variable models.}---
Hidden variable models (HVMs) provide a broad, general approach to considering possible physical explanations underlying observed data\footnote{Sometimes HVMs may also be referred to as ontological models -- attempts to capture an underlying ontology or physical reality. Note however that in some circumstances this terminology may implicitly assume additional structure to what define for HVMs here \cite{Spekkens2005}.}. 
We usually place assumptions on HVMs to reflect notions of classicality---for instance outcome determinism and parameter independence \cite{abramsky2011sheaf}.

Formally a HVM on $\XMO$ is a triple $\tuple{\Lambda, p, (h^\lambda)_{\lambda \in \Lambda}}$ where:
\begin{enumerate*}[label=(\roman*)]
    \item $\Lambda$ is a finite space of hidden variables;
    \item $p$ is a probability distribution on $\Lambda$;
    \item for each $\lambda \in \Lambda$, $h^\lambda$ is a behaviour on $\XMO$.
\end{enumerate*}

Essentially each hidden variable has a behaviour associated with it. We may only have probabilistic information about which hidden variable pertains, reflected by the probability distribution $p$. This gives rise to an effective overall empirical behaviour $h \defeq \sum_{\lambda \in \Lambda} p(\lambda) h^\lambda$. 
A sufficient condition for $h$ to be \textit{non-signalling} is \textit{parameter independence}, which requires that for all $ \lambda \in \Lambda$ and contexts $C_1, C_2 \in \Mc$, $h^\lambda_{C_1}|_{C_1 \cap C_2} = h^\lambda_{C_2}|_{C_1 \cap C_2}$. 

To ensure \textit{sharpness} we can impose \textit{outcome determinism} \ie that $\forall \lambda$, $\forall C \in \Mc$, $h^\lambda_C$ is a Dirac distribution on a given joint outcome in $O_C$.
Note that, in contrast to parameter independence, it is always possible to find a decomposition of an empirical behaviour into deterministic hidden variables (which may be signalling). 
The following proposition is a corollary of \cite[Prop. 3.1 and Th. 8.1]{abramsky2011sheaf}.
\begin{proposition}[from \cite{abramsky2011sheaf}]
\label{prop:sheafequiv}
An empirical behaviour $e$ is noncontextual ($\CF(e)=0$) if and only if it is realisable by a parameter-independent, deterministic HVM.
\end{proposition}

We relax the assumptions of OD and PI below. 
\begin{definition}[$(1-\sigma)$ PI HVM]\label{def:NSHVM}
A hidden-variable model is said to be $(1-\sigma)$ parameter-independent if for all hidden variables $\lambda$, there exists a decomposition of $h^\lambda$ of the form
$
    h^\lambda = (1 - \sigma_\lambda) h^\lambda_\text{NS} + \sigma_\lambda h'^{\lambda}
$
with $h^\lambda_\text{NS}$ a parameter-independent model and with necessarily $\sigma_\lambda \leq \sigma$.
\end{definition}
Note that $\sigma$ is not necessarily the signalling fraction of the overall empirical behaviour realised by the HVM, but rather the maximum parameter dependence allowed for each hidden variable behaviour.

\begin{definition}[$(1-\eta)$ OD HVM] \label{def:ODHVM}
A hidden-variable model is said to be $(1-\eta)$ outcome deterministic if for all hidden variables $\lambda$, there exists a decomposition of $h^\lambda$ of the form
$
    h^\lambda = (1-\eta_\lambda) h^\lambda_\text{OD} + \eta_\lambda h''^{\lambda}
$
with $h^\lambda_\text{OD}$ an outcome deterministic behaviour (which may be parameter-dependent) and $\eta_\lambda \leq \eta$.
\end{definition}

An interesting and well-studied behaviour is the PR-box \cite{popescu1994quantum}.
It is a non-signalling, maximally contextual ($\CF=1$) empirical model.
In other words it does not admit a HVM realisation that satisfies the strict definitions of parameter independence and outcome determinism.
However, under our relaxed definitions, we could in fact explain PR-box correlations by HVMs by allowing either \textit{(i)} a degree of outcome non-determinism or  \textit{(ii)} parameter dependence (or a combination of both).

To see how PR-box behaviour can be explained by outcome non-deterministic HVMs, we
assume parameter independence ($\sigma = 0$) and note that it only admits a $(1-\eta)$ outcome-deterministic HVM with $\eta=0.5$. 
To see this, the model is composed of a single hidden variable behaviour, the PR-box itself, which is parameter-independent:
\begin{equation}
    h_{\text{PR}} = h^\lambda_{\text{PR}} \,.
\end{equation}
Since it's possible to express the hidden variable behaviour of the PR-box as a mixture of two deterministic (but parameter-dependent) behaviours:
\begin{equation}
    h^\lambda_{\text{PR}} = \frac{1}{2} h^\lambda_{M_1} + \frac{1}{2} h^\lambda_{M_2}\,,
    \label{eq:decomposition_PR}
\end{equation}
then, by Definition~\ref{def:ODHVM}, $\eta = \sfrac{1}{2}$.
On the other hand the PR-box could also arise from a purely outcome deterministic ($\eta=0$) HVM. Revisiting the decomposition from Eq.~\eqref{eq:decomposition_PR}, this can be seen as describing the PR-box by an outcome deterministic hidden variable model containing two maximally signalling hidden variable behaviours ($\sigma=1$) which each pertains with equal probability: $e^{\text{PR}} = \frac{1}{2} h^{\lambda_1}_{M_1} + \frac{1}{2} h^{\lambda_2}_{M_2}$. For a longer discussion on the PR-box we refer the reader to appendix (A.d).

Note that the assumptions of determinism and parameter independence are not quite on an equal footing.
If parameter independence is relaxed completely (that is, allowing $\sigma = 1$) then determinism becomes trivial in the sense that any behaviour will admit deterministic HVMs.
On the other hand, if determinism is relaxed, but parameter independence is respected, then only a non-signalling behaviour may be constructed. A related discussion can be found in \cite{norsen2006bell}.

\section{Continuity of the contextual fraction}
\label{sec:continuity}

Since realistic experiments are inherently noisy, it is important that any measure of contextuality is robust to noise.
At the very least it should satisfy some notion of continuity with respect to some total variation distance $V$ generalised to empirical models (see Supplemental Material).
The main theorem in this section is thus significant independently of the rest of the present work,
and adds to the list of desirable properties of the contextual fraction measure identified in \cite{abramsky2017contextual}.

\begin{theorem}
\label{thm:continuity}
Let $e$ and $e'$ be empirical models on the measurement scenario $\XMO$. If $V(e,e') \leq \varepsilon$ for $\varepsilon > 0$ then $\vert \CF(e) - \CF(e') \vert \leq \vert \Mc \vert \varepsilon $.
\end{theorem}

A detailed proof is given in Supplemental Material.

\begin{proof}[Sketch of proof]
We use the theory of finite dimensional linear programming, and more precisely perturbation theory \cite[Section 5.6]{boyd2004convex}.
The idea is to control perturbation of the optimal value of the primal program by exhibiting optimal solutions of the dual program and a perturbed dual program.
\end{proof}

\section{Corrected Bell and Noncontextuality Inequalities}
\label{sec:mainresult}

The nonclassicality criterion $\CF > 0$ applies for perfectly sharp and no-signalling behaviours. Below we want to relax these assumptions and still derive an inequality that captures genuine nonclassicality.

\subsection{Relaxing parameter independence and sharpness}
\label{sec:inequalities}

\begin{lemma}
\label{lemma:corrected}
Let $\tuple{ \{ \lambda \}, \delta_\lambda, (h^\lambda)}$ be a $(1 - \eta_\lambda)$ OD and $(1 - \sigma_\lambda)$ PI HVM.
Then,
if $2\eta_{\lambda} + \sigma_{\lambda} < 1$, the contextual fraction of the hidden variable model satisfies $\CF(h^\lambda) \leq \eta_\lambda$.
\end{lemma}

A detailed proof is given in Supplemental Material.

\begin{proof}[Sketch of proof]
Using Definition~\ref{def:ODHVM}, we can distinguish two cases for an optimal outcome deterministic decomposition: either $h^\lambda_\text{OD}$ is parameter-independent, in which case we compare with \eqref{eq:cf} to see that the bound is respected, or $h^\lambda_\text{OD}$ is parqameter-dependent, in which case we show that $\sigma_\lambda + 2\eta_\lambda \geq 1$ contradicting the initial assumptions.
\end{proof}

\begin{theorem} \label{thm:CF_bound}
Let $e = \sum_\lambda p(\lambda) h^\lambda $ be a behaviour realisable by a $(1-\sigma)$ PI and $(1 - \eta)$ OD HVM $\tuple{\Lambda, p, (h^\lambda)_{\lambda \in \Lambda}}$ such that $\sigma + 2\eta < 1$. Then its contextual fraction is bounded above by: $\CF(e) \leq \eta$.
\end{theorem}

A detailed proof is given in Supplemental Material.

\begin{proof}[Sketch of proof]
By applying Lemma~\ref{lemma:corrected} to all hidden variable behaviours and using the convexity of the contextual fraction, we obtain Theorem~\ref{thm:CF_bound}.
\end{proof}

A similar result was obtained previously in \cite{hall2011Relaxed} specifically for the Bell-CHSH scenario, with similar definitions of parameter dependence and determinism.
In comparing the works, one should note that here we assume measurement independence (also related to experimenter free-will), that is $p(\lambda|x,y) = p(\lambda|x',y') = p(\lambda)$ and thus don't include correction terms related to measurement dependence in our inequality.

The condition $2\eta_{\lambda} + \sigma_{\lambda} < 1$ from Lemma~\ref{lemma:corrected} can be motivated either from experimental considerations or by taking a more theoretical approach. First, in real experimental setups, we expect a low-noise regime so that we remain close to the quantum (or classical) set. In this case the ontological conditions $2\eta_{\lambda} + \sigma_{\lambda} < 1$ should be respected. Secondly, for $n$-cycle scenarios~\cite{araujo2013all}, one can find a hidden variable behaviours which respect $2\eta_{\lambda} + \sigma_{\lambda} = 1$ such that $\CF(h^\lambda) = 1$. The proof is given in the Supplemental Material and Figure~\ref{fig:PR_example} gives a geometric intuition for why this holds.
In this case, the inequality for genuine nonclassicality ($\CF > 1$) would become trivially unsatisfiable. 
For more general scenarios it is an open problem to know whether there always exists at least one hidden variable behaviour such that $\CF = 1$ when $2\eta_\lambda + \sigma_\lambda \geq 1$.

We note that in the case of determinism ($\eta=0$), due to Definitions~\ref{def:NSHVM} and \ref{def:ODHVM}, we have either $\sigma=0$ or $\sigma=1$. Due to the condition $\sigma + 2\eta < 1$, Theorem~\ref{thm:CF_bound} can either be applied for a parameter-dependent HVM or it cannot be applied since $\sigma = 1$.
Instead, this result is of most use when parameter dependence can be accounted for by a lack of determinism: $\eta \geq \sigma$.

\subsection{Relaxing only parameter independence} 
\label{section:PIrelaxed}

As noted earlier, parameter dependence and determinism are not on an equal footing. 
For instance relaxing completely parameter independence trivialises outcome determinism.
We note that our previous derivation in Section~\ref{sec:inequalities} is quite rigid for determinism.
Following Definition~\ref{def:NSHVM}, if we impose outcome determinism then the following holds:
\begin{equation}
    \eta = 0 \implies \sigma \in \{0,1\}\,.
\end{equation}
This implies that if one has a perfectly sharp experiment, then parameter dependence is either forbidden $\sigma = 0$ or always maximally relaxed $\sigma = 1$. 

Let us review another way of looking at only relaxing parameter-independence while preserving outcome determinism.
Another form of relaxation has been used in various references e.g.~\cite{guhne2010compatibility,winter2014does}. These relaxations have the assumption of outcome determinism ($\eta=0$), defined in the same manner as Definition~\ref{def:ODHVM} but they relate parameter dependence to another quantity that we will call $\sigma'$ here. Generally speaking, in the case of determinism, each hidden variable behaviour $h^\lambda$ is either parameter-independent or maximally parameter-dependent. Thus we can split the sum of all hidden variables depending on their parameter-dependence:
\begin{equation}
    e = \sum_{h^\lambda \in \text{PI}} p(\lambda) h^\lambda + \sum_{h^\lambda \notin \text{PI}} p(\lambda) h^\lambda\,,
\end{equation}
where PI here is used as a shorthand for the set of hidden variable behaviours that are parameter-independent.

The new definition of parameter independence is now a bound on the probability to have parameter-dependent hidden variables by $\sigma'$:
\begin{equation}
    \sum_{h^\lambda \notin \text{PI}} p(\lambda) \leq \sigma' \Mdot
    \label{eq:bound_signalling_distrib_epsilon}
\end{equation}
This in turn leads easily to a bound on the contextual fraction (due to its convexity):
\begin{equation}  \label{eq:CF_bound_epsilon_relation_work}
    \begin{aligned}
        \CF(e) &\leq \sum_{h^\lambda \in \text{PI}} p(\lambda) \CF(h^\lambda) + \sum_{h^\lambda \notin \text{PI}} p(\lambda) \CF(h^\lambda)\\
        \CF(e) &\leq \sum_{h^\lambda \notin \text{PI}} p(\lambda) \CF(h^\lambda) \\
        \CF(e) &\leq \sum_{h^\lambda \notin \text{PI}} p(\lambda) \\
        \CF(e) &\leq \sigma' \,.
    \end{aligned}
\end{equation}
Note again that this derivation only holds for deterministic hidden variables.

\section{Approaches to experimental bounds}
\label{sec:physical}

To relate Lemma~\ref{lemma:corrected} to statistics gathered during experiments, it is necessary to either physically impose restrictions on the values of $\sigma$ or $\eta$ or to make assumptions on their bounds based on observations from experiments.
In Bell scenarios, for example, separation of parties allows us to make the assumption that the HVM should be parameter-independent \ie $\sigma = 0$.
In this case, Fine's theorem \cite{Fine1982} (more precisely its generalisation in \cite{abramsky2011sheaf}) states that outcome determinism ($\eta =0$) is equivalent to non-contextuality ($\CF = 0$).
However in the absence of spacelike separation ($\sigma >0$), we need to make assumptions on how experimental data is related to possible HVMs.

\paragraph{Parameter-independence.}
By definition, $\sigma \geq \SF(e)$ for any HVM realisation of a behaviour $e$. It means that the allowed parameter-dependent fraction $\sigma$ cannot be lower than the observed signalling fraction of the behaviour $e$ since $\SF$ is the optimal solution of a minimisation program.
In a general manner, one wishes however to be able to upper bound the parameter dependence of the hidden variables according to the observed signalling:
\begin{equation*}
    \sigma \leq f\left(\SF(e) \right)\,,
\end{equation*}
where the function $f$ is to be chosen, perhaps motivated by experimental considerations, through partial device characterisation, space-like shielding, or any other appropriate method.
The simplest assumption would thus be $\sigma = \SF(e)$ but it is important to recall that whereas parameter independence of a HVM implies that the empirical behaviour it realises is non-signalling, the converse does not hold in general, so this is a strong assumption to impose.
Note, however, that the analogous assumption is made in the contextuality-by-default approach to contextuality which also applies in the presence of signalling,
and has been employed in experimental tests of contextuality analysed via contextuality-by-default \cite{kujala2015necessary} and the present approach \cite{fyrillas_certified_2023}.

\paragraph{Outcome determinism}
For determinism, an analogous approach would be to bound $\eta$ by an appropriate quantification of the unsharpness of the empirical measurements.
A convenient approach could be to follow the definition of sharpness found in \cite{chiribella2014measurement}.
Each context of measurements could be performed not once, but twice in quick succession,
and the probabilities of seeing a change in outcomes between successive measurements would give an indication of the unsharpness.
This does however entail an additional assumption of measurement \emph{non-invasiveness} \cite{clemente2015necessary,moreira2019quantifying}, \ie that measurement does not alter the underlying state of the system (neither the value of $\lambda$ nor the response $h^\lambda$).
In this way it allows for effectively accessing properties of $h^\lambda$. 
This can be a sensible approach in experiments such as that of Kirchmair \emph{et al.}\ \cite{kirchmair2009state,guhne2010compatibility} in which sequences of measurements were performed on trapped ions.

For cases where repeated measurements of a system are not practical, for example in optical setups where photons are absorbed during the measurement, we must follow a different approach. 
One possibility is to assume the existence of a particular set of preparations that allows us to assess if a measurement device is responding deterministically or not (similar to the assumption of a tomographically complete set of preparations used in \cite{spekkens2009preparation,spekkens2014status,kunjwal2018statistical,kunjwal2019beyond}).
This is motivated by the fact that in the ideal quantum mechanical description for any sharp measurement there is always a set of states which return deterministic responses.
If we assume that we have such special, preparable states, they can be used to quantify outcome determinism.

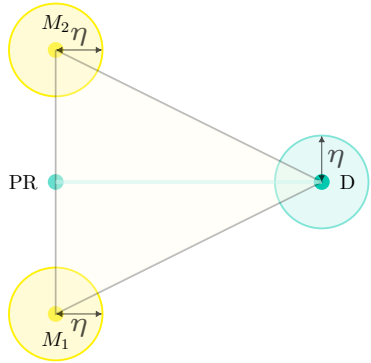
\begin{figure}
\centering
    \scalebox{1.75}{\makeatletter
\tikzoption{canvas is plane}[]{\@setOxy#1}
\def\@setOxy O(#1,#2,#3)x(#4,#5,#6)y(#7,#8,#9)%
  {\def\tikz@plane@origin{\pgfpointxyz{#1}{#2}{#3}}%
   \def\tikz@plane@x{\pgfpointxyz{#4}{#5}{#6}}%
   \def\tikz@plane@y{\pgfpointxyz{#7}{#8}{#9}}%
   \tikz@canvas@is@plane
  }
\makeatother  

\begin{tikzpicture}[scale=1.]

\tikzstyle{vertex}=[circle, draw, inner sep=0pt, minimum size=3pt]
\newcommand{\vertex}{\node[vertex]}

\draw[yellow!100, fill=yellow!20, opacity=1] (0,1) circle (10pt);
\draw[yellow!100, fill=yellow!20, opacity=1] (0,-1) circle (10pt);
\draw[quantumturquoise!100, fill=quantumturquoise!20, opacity=.5] (2,0) circle (10pt);

\vertex[quantumturquoise!100, fill=quantumturquoise!100] (p1) at (2,0) {};

\vertex[quantumturquoise!60, fill=quantumturquoise!60] (p2) at (0,0) {};

\vertex[yellow!80, fill=yellow!80] (p3) at (0,1) {};
\vertex[yellow!80, fill=yellow!80] (p4) at (0,-1) {};

\node[above] (p0) at (0,1) { \scalebox{.4}{$M_2$} };
\node[below] (p1) at (0,-1) { \scalebox{.4}{$M_1$}};
\node[left] (p2) at (0,0) { \scalebox{.4}{PR} };
\node[right] (o3) at (2,0) { \scalebox{.4}{D}};

\draw[opacity=.3, fill=none]  (0,1) -- (0,-1) -- (2,0) -- (0,1);

\draw[draw=none,fill=yellow!10,opacity=0.3] (0,1) -- (0,-1) -- (2,0) -- (0,1);

\draw[quantumturquoise!20,fill=quantumturquoise!20,opacity=.5,line width=0.25mm] (2,0) -- (0,0) -- (2,0);

\draw[{Latex[length=0.5mm,width=0.5mm]}-{Latex[length=0.5mm,width=0.5mm]}, very thin, opacity=0.7] (0,1) -- (10pt,1);
\draw[{Latex[length=0.5mm,width=0.5mm]}-{Latex[length=0.5mm,width=0.5mm]}, very thin, opacity=0.7] (0,-1) -- (10pt,-1);
\draw[{Latex[length=0.5mm,width=0.5mm]}-{Latex[length=0.5mm,width=0.5mm]}, very thin, opacity=0.7] (2,0) -- (2,10pt);

\node[above, opacity=0.7] (e) at (5pt,0.9) { \tiny $\eta$};
\node[below, opacity=0.7] (e2) at (5pt,-0.9) { \tiny $\eta$};
\node[right, opacity=0.7] (e3) at (1.9,5pt) { \tiny $\eta$};

\end{tikzpicture}}
    \caption{A schematic diagram of a vertical cut of Figure~\ref{fig:Spolytope}. 
    Here, discs of radius $\eta$ have been added around the deterministic points to represent the accessible hidden variable behaviours for a given $\eta$. Theorem~\ref{thm:CF_bound} applied in this scenario means that whenever $\sigma + 2\eta < 1$, the only accessible behaviours are the ones within the light blue disc surrounding the deterministic classical blue point. On the other hand, when $\sigma + 2\eta \geq 1$ the behaviours inside the intersections of the yellow discs (centred on maximally signalling vertices) and the signalling polytope become accessible, and thus 
    $\CF = 1$ is reachable.
    }
    \label{fig:PR_example}
\end{figure}

\section{Relationship to previous results}

We now discuss how our results relate to others that have previously appeared in the literature \cite{guhne2010compatibility,winter2014does,lapkiewicz2011experimental,hu_experimental_2016,marques2014experimental}.
To this end we will consider how to fine-tune the parameters appearing in Theorem~\ref{thm:CF_bound} or the derivation in Section~\ref{section:PIrelaxed} to adjust it to some specific experimental setups or data. 
In particular we will show how several previous results are contained and can be extended within our approach.
In this section our intention is to be illustrative rather than exhaustive with respect to existing literature.

\subsection{Relaxing determinism in known frameworks}
As already discussed, in the case of determinism, we could apply directly the derivation from Section~\ref{section:PIrelaxed}. 
In this case we would obtain directly the results from \cite{guhne2010compatibility,winter2014does}.
However, one may also be interested in relaxing determinism and looking at  how our result obtained in Theorem~\ref{thm:CF_bound} could be apply in this setting.  
In the following, we thus aim to relate Theorem~\ref{thm:CF_bound} to results imposing determinism.

\hspace{10pt}\textbf{Ref.~\cite{guhne2010compatibility}.}
We will show that relaxing determinism is an alternative to relaxing parameter dependence for a high enough relaxation.
Reference \cite{guhne2010compatibility} is concerned with the order in which measurements are performed within a context but this is not something we have sought to capture in our approach. 
Our approach may still be applied, but will simply treat such orderings as the same context.
The correction term in~\cite{guhne2010compatibility} arises as a sum over probabilities that the outcome of an observable `flips' in a given context. Roughly:
\begin{equation*}
    \varepsilon = \sum_{C \in \mathcal{M}} p^\text{flip}[C]\,,
\end{equation*}
where $p^\text{flip}[C]$ is the probability that at least one measurement outcome in context $C$ is flipping. As an example, in the CHSH scenario where $C = AB$ we have $p^\text{flip}[AB] = p\left [ (a = 1|A) \text{ and } (a = 0|AB)\right ] + p\left [ (a = 0|A) \text{ and } (a = 1|AB)\right ]$ where $a$ is the outcome of $A$.
In the CHSH scenario, the bound becomes~\footnote{Note that this is an upper bound, since it could be that the flipping of $A$ and $B$ always occur together, which means we would double the impact of the flipping on the violation.}:
\begin{equation} \label{eq:CF_less_sum_pflip}
\begin{aligned}
    \CF(e) &\leq p^\text{flip}[AB] + p^\text{flip}[AB'] + p^\text{flip}[A'B] + p^\text{flip}[A'B'] \\
    &\leq \varepsilon \,,
\end{aligned}
\end{equation}
We propose instead to consider the case where the flipping occurs only due to lack of determinism. We thus relax determinism by making the assumption that signalling observed at the empirical level is the result of non-determinism only:
\begin{equation}
    \eta \defeq \max_{C\in\mathcal{M}} \left ( p^{\text{flip}}[C]\right )\,.
\end{equation}
In this case, we are in the regime where $\eta \geq \sigma$. If $2\eta + \sigma < 1$, then we can simply apply Theorem~\ref{thm:CF_bound}:

\begin{equation}
    \CF(e) \leq \max_{C\in\mathcal{M}}\left ( p^\text{flip}[C] \right ) \,.
\end{equation}

\hspace{10pt}\textbf{Ref.~\cite{winter2014does}.} This work also aims at correcting inequalities with an additional term depending on
\begin{equation}
    \label{eq:relaxation_winter2014}
    \text{Pr}(o_x^{C} \neq o_x^{C'}) \leq \varepsilon
\end{equation}
where $o_x^{C}$ is the outcome of some observable $x$ in context $C$, and $C\cap C' \neq \emptyset$.
The corrected bound proposed by~\cite{winter2014does} is the following:
\begin{equation}\label{eq:winter}
    \CF(e) \leq \frac{\sum_i \omega_i (k_i - 1)}{\beta_{max} - \beta_{cl}} \varepsilon \,,
\end{equation}
where $\omega_i$ are the weights of the inequality being tested, $\beta_{max}$ is its algebraic bound and $\beta_{cl}$ its classical bound, and $k_i$ is the degree of measurement $i$ (the number of times it appears in different contexts, e.g.\ $k_i = 2$ for all measurements in the CHSH scenario).
In the same spirit as our approach to Ref.~\cite{guhne2010compatibility}, we assume that the difference of the outcomes in different context is just due to lack of determinism:
\begin{equation}
    \label{eq:relaxation_winter2014_modified}
    \text{Pr}(o_x^{C} \neq o_x^{C'}) \leq \eta \, .
\end{equation}
Which, by application of Theorem~\ref{thm:CF_bound}, gives:
\begin{equation}
    \CF(e) \leq \eta\,.
\end{equation}
We now compare the two bounds in different known scenarios.

Following~\cite{winter2014does}, we are interested in ranges of $\varepsilon$ for which this bound on the contextual fraction is less than or equal to $\CF_{Q}$, the maximum attainable contextual fraction over quantum mechanical models.
Otherwise noise would be too high to tell apart quantum behaviours from noisy classical ones. 
While there is no general relation between $\sum_i \omega_i (k_i - 1)$ and $\beta_{max} - \beta_{cl}$ in Equation~\eqref{eq:winter}, in practical examples note that the numerator is much larger than the denominator.
We can take as an example the Peres-Mermin square~\cite{mermin_hidden_1993,kirchmair2009state}.
In our approach saturating the quantum bound, which for the Peres-Mermin square coincides with the algebraic bound, would require $\eta = 1$,
whereas for \cite{winter2014does} it is saturated once $\varepsilon = \frac{6 - 5}{72} \approx 0.0138$.
Thus for the same ontological assumptions, we obtain a non-classicality witness which remains useful in a much higher noise regime.

\subsection{Application to experimental results}

This subsection considers experimental works that did not apply corrected inequalities, or to which we wish to directly compare the result of Theorem~\ref{thm:CF_bound}. Unlike in the previous section, we directly take the values from the experiments to compute the corrected bound. This subsection may serve as an example of various means to compute $\eta$ and $\sigma$ from experiments, by making assumptions on the hidden variables from the empirical behaviour.

\hspace{10pt}\textbf{Ref.~\cite{marques2014experimental}.} The authors reproduce a Hardy-like KCBS inequality, such that in theory $p(1,1|i,i+1) = 0$ for  $i\in\{1,2,3,4,5\}$ for sharp measurements. Yet in practice they do obtain some events $(1,1|i,i+1)$, which means that part of the process is noisy (whether it is preparation, measurement or transformation). It makes sense to assume then that measurements are unsharp and the sharpness of measurement can be a good indicator of the hidden variable properties. If the detector fails $\eta$ of the time then the HV behaviours would be $1-\eta$ of the time deterministic (same detectors click) and $\eta$ of the time giving a random result.
Thus we propose to chose
\begin{equation}
    \eta = \max_i\left (p(1,1|i,i+1)\right)
\end{equation}
such that the unsharpness reflects the value of $\eta$. With this choice we retrieve the same result as that of \cite{marques2014experimental}. 

The assumption made in the aforementionned reference is that $\varepsilon$ of the time the maximum violation is attained, leading to the following corrected inequality :
\begin{equation}
    \left|\braket{KCBS} \right| \leq (1 - \varepsilon) \beta_{cl} + \varepsilon \beta_{max}\,,
\end{equation}
which in terms of $\CF$ is equivalent to
\begin{equation}
    \CF \leq \varepsilon\,.
\end{equation}
Our result---derived from ontological considerations---
states that the maximum non-determinism is given by $\eta$, leading to the same mathematical conclusion:
\begin{equation}
    \CF \leq \eta \Mdot
\end{equation}
Since we arrive at the same mathematical conclusion, we can tell that their violation is robust to the amount of observed unsharpness. They get $\eta = 0.021$ while they obtain a violation  $\CF \approx 0.16$ which is largely above the expected bound given by Theorem~\ref{thm:CF_bound}.

\hspace{10pt}\textbf{Ref.~\cite{lapkiewicz2011experimental}.} Our work may be applied to the experimental procedure described in \cite{lapkiewicz2011experimental} to test the KCBS inequality~\cite{klyachko_simple_2008}.
The experimental setup is designed to ensure that measurements cannot have context dependence, up to a set of assumptions, except for one measurement $A_1$ appearing in contexts $A_1 A_2$ and $A_5 A_1$.
They thus label the measurement appearing in these contexts as $A_1$ and $A_1'$ respectively, so that the modified KCBS inequality looks like:
\begin{equation}
    \braket{A_1 A_2} + \braket{A_2 A_3} + \braket{A_3 A_4}
    + \braket{A_4 A_5} + \braket{A_5 A_1'} \geq - 3 - \kappa\,,
\end{equation}
where $\kappa$ is a correction term depending on the correlation between $A_1$ and $A_1'$.
In our approach, rather than considering these as distinct measurements, we would rather account for any discrepancy as unsharpness and attribute this to either parameter dependence or outcome indeterminism at the hidden variable level.
As $A_1$ is a dichotomic measurement this amounts to attributing $\eta$ in the following way:
\begin{equation}
    \eta = \max_{o\in \{-1,1\} }p(A_1 = o | A_1' = -o) \Mdot
\end{equation}
The bound of \cite{lapkiewicz2011experimental} is tighter: they obtain $\CF \leq 0.041(2)$ versus $\CF \leq 0.072$ as a criterion for classicality when recast in our terms.
A contextuality-by-default analysis led to a normalised bound of $\CF \leq 0.234$ \cite{kujala2015necessary}.
In contrast to both analyses we point to the fact that our approach provides a clear relationship with assumptions at the ontological level.

\hspace{10pt}\textbf{Ref.~\cite{hu_experimental_2016}.} This paper is a photonic implementation of an experiment proposed in \cite{cabello2011Proposal} which aims at closing the compatibility loophole. It uses photonic qutrit in order to violate the following inequality:
\begin{equation}
    P(D_1^A = 1 | D_0^B = 1) - P(T_0^A = a_0 | D_0^B = 1) - P(T_1^A = a_1 | D_0^B = 1) \leq 0,
\end{equation}
where $D_i^O$ and $T_i^O$ are observables and $a_0$ and $a_1$ are further parameters that describe the measurements.
Here we wish to test our results to compare the experimental results to the theoretical bound. In this paper we can take departure from the expected quantum behaviour as unsharpness at the level of the probabilities, such that:
\begin{equation}
    \eta = \max_{x}(|p_\text{th}(x) - p_\text{exp}(x)|)\,.
\end{equation}
where $x$ is taken in the set of all possible events: $D_1^A = 1 | D_0^B = 1$, $T_0^A = a_0 | D_0^B = 1$ and $T_1^A = a_1 | D_0^B = 1$. $p_\text{th}$ and $p_\text{exp}$ are the theoretical probability and the experimental probability respectively.
The highest difference is found for $p_\text{exp}(T_1^A = a_1 | D_0^B = 1)$, where the theoretical prediction is $0$ and the experimental result is $0.010 \pm 0.001$. Then we can assume that $\eta \leq 0.01$. On the other hand, signalling is measured to be of the order of $10^{-3}$ thus we can safely apply Theorem~\ref{thm:CF_bound} by assuming that $\sigma = 0.001$:
\begin{equation}
    \begin{split}
        \CF &\leq \eta \leq 0.01,
    \end{split}
\end{equation}
which we can directly compare to their result $\CF = 0.89 > 0.01$ implying that they are indeed witnessing noise-robust contextuality.

\hspace{10pt}\textbf{Ref.~\cite{wang2022Significant}.} This reference is another significant experiment on contextuality which closes the sharpness, detection and compatibility loopholes, thus claiming to be significantly loophole free. The experiment focuses on an inequality equivalent to the CHSH inequality with two ions of different nature in order to avoid interference on the measurements and operations:
\begin{equation}
    \mathcal{C} = \braket{\hat{O}_0\hat{O}_1} + \braket{\hat{O}_1\hat{O}_2} + \braket{\hat{O}_2\hat{O}_3} - \braket{\hat{O}_3\hat{O}_0} \leq 2.
\end{equation}

In order to close the sharpness loophole, they measure the repeatability $R_i$ of the observables $\hat{O}_i$. It corresponds to the number of times the observable $\hat{O}_i$ gives different outcomes when measured twice in a row. The lowest repeatability is found to be $R_0 \approx 97\%$ for $\hat{O}_0$. 
Taking the repeatability as a good indicator of unsharpness, if we consider that both parties have at least a repeatability $R_i \geq 1 - \epsilon = 0.97$ then one can define $\eta = 1 - (1-\epsilon)^2 = 2\epsilon - \epsilon^2$. Thus, $\eta \approx 0.06$.

We can assume that signalling is so small that it can be neglected. In fact, they measure the cross-talk between the two ion species, and they observe that both have a maximum population transfer between $1.9 \times 10^{-6}$ and $4.3 \times 10^{-6}$ when they are excited by the wrong laser which is negligible for their experiment, and we do not consider it here.

We can compute our bound since $2\eta + \sigma \approx 0.12 + 0 < 1$ which satisfies the condition for Theorem~\ref{thm:CF_bound}. Then for the empirical model $e_\mathcal{C}$ associated to $\mathcal{C}$ we have:
\begin{equation}
    \CF(e_\mathcal{C}) \leq 0.06,
\end{equation}
which translates in term of the inequality to
\begin{equation}
    \begin{split}
        \mathcal{C} &\leq 2 + (4 - 2) \eta \\
        &\leq 2.12.
    \end{split}
\end{equation}
Since they obtain a violation $\mathcal{C} = 2.526 \pm 0.016$, our framework with these assumptions certifies this experiment as contextual.

\section{Conclusion}

In this work, we present a new bound on the contextual fraction allowed by hidden-variable models that have access to limited parameter dependence (related to empirical signalling) and nondeterminism (related to empirical unsharpness).
We considered several approaches to bounding parameter dependence and nondeterminism of an HVM, which inevitably  employ additional assumptions that relate these HV properties to empirical signalling and unsharpness.
This allows for a robust testing of non-classicality for noisy measurements. 

The main goal of this work is to achieve the above in the broadest and simplest way.
It is not our aim to assert which assumptions are valid, though this is a crucial question, but rather to provide clear mathematical results and bounds which are straightforward to check, around which such discussions may take place. 
A benefit of our approach is that it allows clarity in where the additional assumptions must sit---i.e. in experimentally accessing the values for $\eta$ and $\sigma$.
Note that no such assumptions are brought into play yet in proving the bounds of Lemma~\ref{thm:CF_bound} and Theorem~\ref{lemma:corrected};
these follow from the definitions.
We presented a variety of possible assumptions that would allow $\eta$ and $\sigma$ to be empirically bounded by looking at previous works in the literature \cite{winter2014does,guhne2010compatibility,lapkiewicz2011experimental,hu_experimental_2016,marques2014experimental}.
In this way, we see that our framework is broad enough to subsume a wide variety of previous approaches.

With respect to applications, and beyond purely foundational considerations, it has also recently been demonstrated that bounds on $\eta$ and $\sigma$ can be translated into constraints on adversaries or computational models, for example in generating certified randomness in the presence of bounded signalling on a microchip \cite{fyrillas_certified_2023}.
Seeking clarity and flexibility in the role and choice of the assumptions that can allow empirical access to parameter dependence and nondeterminism can thus be of benefit for future applications of these inequalities to quantum information and computing more broadly.

\section*{Acknowledgements}
The notion of the signalling fraction originally emerged in unpublished work by Samson Abramsky, Rui Soares Barbosa and SM. The authors thank the participants of the 2021 and 2022 editions of the Quantum Contextuality in Quantum Mechanics and Beyond Workshop for their feedback on earlier versions of this work. \\[5pt]
This work has received funding from the PEPR integrated project EPiQ ANR-22-PETQ-0007 part of Plan France 2030, and from BPI France Concours Innovation PIA3 projects DOS0148634/00 and DOS0148633/00 -- Reconfigurable Optical Quantum Computing.

\section*{Author contibution}

\noindent K.V. conceptualization, formal analysis (section 4, 5, 6), investigation, visualization, writing – original draft; 
P-E.E: conceptualization, formal analysis (section 3, 5), investigation, supervision (section 4, 6), visualization, writing – original draft; 
B.B: validation, writing: review \& editing; 
A.S.: Formal analysis, Validation;S.M. conceptualisation, formal analysis, investigation, supervision, writing: review \& editing;
D.M. conceptualisation, formal analysis, investigation, supervision, writing: review \& editing;
S.M. and D.M. contributed equally to this work.

\bibliographystyle{RS}
\bibliography{biblio}

\newpage

\begin{appendices}

\renewcommand{\theequation}{\thesection.\arabic{equation}}


\section{General framework for contextuality}

This section summarises some of the main ideas from \cite{abramsky2011sheaf,abramsky2017contextual},
setting out the definition of contextuality, assumptions on hidden variable models, and how these are related, and finally defining the \textit{contextual fraction}, a quantitative measure of the degree to which observed empirical data is contextual.

\subsection{Scenarios and Behaviours}
\label{sec:app_measurementscenario}

An abstract description of a particular experimental setup is formalised as \textit{a measurement scenario}. 
\begin{definition}[Measurement scenario]
A measurement scenario is a triple $\XMO$ where:
\begin{itemize}
    \item $X$ is a finite set of measurement labels.
    \item $\Mc$ is a covering family of $X$ \ie it is a set of subsets of $X$ such that $\bigcup_{C \in \Mc} C = X$. 
    The element $C \in \Mc$ are taken as maximal contexts and represent maximal sets of compatible observables \footnote{From recent works using the Sheaf-theoretic treatments for contextuality \cite{barbosa2014monogamy,barbosa2015contextuality,caru2018towards,karvonen2018categories,abramsky2019comonadic}, $\Mc$ is often taken as the set of maximal contexts requiring that $\Mc$ be an anti-chain with respect to subset inclusion \ie that no element of the family is a proper subset of another.}. 
    \item $O = (O_x)_{x \in X}$ is a finite set of outcomes for each measurement. If some set of measurements $U \subseteq X$ are considered together then the corresponding joint outcome set is given by the (Cartesian) product of the respective outcome spaces: $O_U = \prod_{x \in U} O_x$. 
\end{itemize}
\end{definition}
For example, the setup corresponding to the usual CHSH experiment \cite{CHSH1969} or (2,2,2) Bell scenario---two spacelike separated parties, Alice and Bob, each have two possible binary measurements---is described by:
\begin{equation}
    \label{eq:CHSH_measurement_scenario}
    \begin{split}
        & X=\enset{a,a',b,b'},\\ 
        & \Mc = \enset{ \, \enset{a,b}, \, \enset{a,b'}, \, \enset{a',b}, \, \enset{a',b'} \, }, \\
        & \forall x \in X, O_x = \left\{0,1\right\} .
    \end{split}
\end{equation}

Next, given the description of the experimental setup, either calculating theoretical predictions for admissible joint outcomes or performing repeated runs of the experiment with varying choices of measurement context and recording the frequencies of the corresponding joint events results in a probability table which is formalised as an \textit{empirical model}. 

\begin{definition}[Empirical model/behaviour]
Given a measurement scenario $\XMO$, an empirical model (or behaviour) is a family $e = (e_C)_{C \in \Mc}$ where $e_C$ is a probability distribution on the joint outcome space $O_C$.
\end{definition}

For example, an empirical model corresponding to the usual CHSH experiment is given by Table \ref{tab:empCHSH}.
\begin{table}[!ht]
	\centering
	\begin{tabular}{cc||cccc}
		 A & B & $00$ & $01$ & $10$ & $11$ \\ \hline
		 $a$ & $b$ & $p_1$ & $p_2$ & $p_2$ & $p_1$ \\
		 $a$ & $b'$ & $p_1$ & $p_2$ & $p_2$& $p_1$ \\
		 $a'$ & $b$ & $p_1$ & $p_2$ & $p_2$ & $p_1$ \\
		 $a'$ & $b'$ & $p_2$ & $p_1$ & $p_1$ & $p_2$
	\end{tabular}
	\caption{Quantum empirical model on the $(2,2,2)$ Bell scenario specifying the probabilities of the joint outcomes for the CHSH model with $p_1 = \frac{2+\sqrt 2}{8}$ and $p_2 = \frac{2-\sqrt 2}{8}$. Here $a_i$ and $b_i$ for $i=1,2$ represent quantum observables for CHSH with eigenvalues relabelled as $0$ and $1$. Joint probabilities are obtained by the Born rule.}
	\label{tab:empCHSH}
\end{table}
We may further require that the empirical data is non-signalling by imposing compatibility conditions across different contexts, that is, for $C,C' \in \Mc$, $e_C|_{C \cap C'} = e_{C'}|_{C \cap C'}$.

\subsection{Contextuality and its quantification via the contextual fraction}
Now that we have defined what we mean by a measurement scenario and an empirical model we can define precisely what we mean by \textit{contextuality} which is a property of the empirical data. 

\begin{definition}[Noncontextuality]
A {non-signalling} empirical model $e$ on a measurement scenario $\XMO$ is said to be noncontextual if there exists a global probability distribution $d$ on $O_X$ such that $e_C = d|_C$.
\end{definition}

This corresponds to a property of extendability for the empirical data \ie the fact that we can extend all probability distributions into a global one from which it is possible to marginalise in order to retrieve a probability distribution for a given context. If such a global distribution cannot be found then $e$ is said to be \textit{contextual}. 

A more refined question than asking if an empirical model $e$ is contextual is to ask what fraction of the empirical data admits a noncontextual explanation. This is formalised as the \textit{contextual fraction} \cite{abramsky2011sheaf,abramsky2017contextual}. Instead of asking for a global probability distribution from which the empirical behaviour $e$ can be deduced by marginalisation at each context, the idea is to look for a subprobability distribution $b$ (\ie that sums to less than 1) on $O_X$ explaining some fraction of the data \ie we require that $\forall C \in \Mc, b|_C \leq e_C$. Ideally, we would like to find such a $b$ maximising $\bm 1 \cdot b = \sum_{i = 1}^{\vert O_X \vert} b_i$.

\begin{definition}[Noncontextual fraction]
Given an empirical model $e$ on a measurement scenario $\XMO$, its noncontextual fraction is given by:
\begin{equation*}
	\NCF(e) \defeq \sup \setdef{\bm 1 \cdot b}{
    b\geq 0 \text{ s.t. } \forall C \in \Mc, b|_C \leq e_C}	\,.
\end{equation*}
\label{def:noncont}
\end{definition}
Note that by construction, we have that $\NCF(e) \in \left[0,1\right]$. The case $\NCF(e)=1$ for non-signalling date corresponds to noncontextuality while $\NCF(e)=0$ corresponds to maximal contextuality or strong contextuality \cite{abramsky2011sheaf}. 
Naturally, the contextual fraction is defined as: $\CF(e) \defeq 1 - \NCF(e)$. We would like to stress that for non-signalling data contextuality corresponds to the case that $\CF(e) > 0$, though our aim is to modify this inequality to include about genuine contextuality of noisy data.

Now we would like to give a more geometric take on the contextual fraction. Given two empirical models $e_1$ and $e_2$ on the same measurement scenario $\XMO$, we can define a third valid empirical by taking a convex sum of $e_1$ and $e_2$: for instance $\lambda e_1 + (1-\lambda) e_2$ for $\lambda \in \left[0,1\right]$ is another valid empirical model. Asking what fraction of a non-signalling empirical model $e$ admits a noncontextual explanation is asking for a convex decomposition of the form:
\begin{equation}
	e = \lambda e^\text{NC} + (1-\lambda) e'\,
	\label{eq:emp_convex_decomp}
\end{equation}
with weight $\lambda$ on the noncontextual part. 
Maximising $\lambda$ yields the noncontextual fraction $\NCF(e)$ (see Figure~\ref{fig:polytopes}).

A convenient property of the contextual fraction is that its computation can be cast as a linear program \cite{abramsky2017contextual}. 
Fix $n \defeq |O_X|$ the number of different global assignments and $m \defeq \sum_{C \in \Mc} |O_C| $ the number of total local assignments ranging over contexts. We use bold notation for vectors. Local assignments can be listed as: $\left\{ \langle C,\bm s \rangle \text{ s.t. } C \in \Mc \text{ and } \bm s \in O_C \right\}$. Then the incidence matrix $\Mrm$ which records the possible restrictions from global to local assignments is a $m \times n$ (0,1)-matrix defined as:
\begin{equation}
    \Mrm[\langle C,s \rangle,g] \defeq 
    \begin{cases}
    1 \text{ if } \bm g|_C = \bm s \\
    0 \text{ otherwise} \Mdot
    \end{cases}
    \label{eq:incidencematSupp}
\end{equation}
To understand its action, read $\Mrm$ column-by-column by fixing a global section $\bm g \in O_X$. For this specific column, $\Mrm$ assigns the value $1$ in the row corresponding to the local section $\bm s$ in context $C$ whenever $\bm g|_C = \bm s$ (and $0$ otherwise).
The empirical data $e$ can be represented as a vector $ \bm \vrm^e  \in \left[0,1\right]^m$ where for a given context $C \in \Mc$ and a local assignment $\bm s \in \Oc_C$, $\bm \vrm^e[\langle C, \, \bm s \rangle] = e_C(\bm s)$. This is the flattened version of the tables (see for instance Table~\ref{tab:empCHSH}) that are usually used to represent empirical models. In the noise-free regime $e$ is noncontextual if and only if there exists a global probability distribution $\bm \drm \in \left[0,1\right]^n$ such that $\Mrm \bm \drm = \bm \vrm^e$ and that $\NCF(e) < 1$ corresponds to the non-existence of such a global distribution $\drm$.
Definition \ref{def:noncont} can be expanded as:

\leqnomode
\begin{flalign*}
    \label{prog:LP-CFSupp}
    \tag*{(P-NCF)}
    \hspace{4cm}\left\{
    \begin{aligned}
        & \quad \text{Find } \bm \brm \in \R^n \\
        & \quad \text{maximising } \bm 1.\bm \brm \\
        & \quad \text{subject to:}  \\
        & \hspace{1cm} \begin{aligned}
            & \Mrm \bm \brm \leq \bm \vrm^e \\
            & \bm \brm \geq \bm 0 \Mdot
        \end{aligned}
    \end{aligned}
    \right. &&
\end{flalign*}
\reqnomode
Interestingly, the dual of this program (with a clever change of variable) is of particular interest since it allows to compute the expression of a Bell inequality optimised to the empirical data \cite{abramsky2017contextual,barbosa2019continuous}. 
We present the dual program of \refprog{LP-CF} without the change of variable $a \defeq \vert \Mc \vert^{-1} \bm 1 - \bm y$ below as it is enough to prove the continuity result of interest.
\leqnomode
\begin{flalign*}
    \label{prog:DLP-CF}
    \tag*{(D-NCF)}
    \hspace{4cm}\left\{
    \begin{aligned}
        & \quad \text{Find } \bm \yrm \in \R^m \\
        & \quad \text{minimising } \bm \yrm.\bm \vrm^e \\
        & \quad \text{subject to:}  \\
        &  \hspace{1cm} \begin{aligned}
            & \Mrm^T \bm \yrm \geq \bm 1 \\
            & \bm \yrm \geq \bm 0 \Mdot
        \end{aligned}
    \end{aligned}
    \right. &&
\end{flalign*}
\reqnomode

\subsection{Quantifying (non-)signalling}
\begin{figure}[ht]
\centering
\begin{minipage}[c]{.45\textwidth}
    \centering
    \scalebox{1}{\begin{tikzpicture}[scale=1]

\tikzstyle{vertex}=[circle, draw, inner sep=0pt, minimum size=3pt]
\newcommand{\vertex}{\node[vertex]}

\draw[quantumturquoise!20,fill=quantumturquoise!20] (0,4) -- (2,5.25) -- (4,4) -- (5.25,2) -- (4,0) -- (2,-1.25) -- (0,0) -- (-1.25,2) -- (0,4);

\draw[quantumturquoise!40,fill=quantumturquoise!40]  (0,0) rectangle (4,4);

\vertex[quantumturquoise!100,fill=quantumturquoise!100] (p1) at (0,0) {};
\vertex[quantumturquoise!100,fill=quantumturquoise!100] (p2) at (0,4) {};
\vertex[quantumturquoise!100,fill=quantumturquoise!100] (p3) at (4,0) {};
\vertex[quantumturquoise!100,fill=quantumturquoise!100] (p4) at (4,4) {};

\vertex[quantumturquoise!60,fill=quantumturquoise!60] (p5) at (2,5.25) {};
\vertex[quantumturquoise!60,fill=quantumturquoise!60] (p6) at (5.25,2) {};
\vertex[quantumturquoise!60,fill=quantumturquoise!60] (p7) at (2,-1.25) {};
\vertex[quantumturquoise!60,fill=quantumturquoise!60] (p8) at (-1.25,2) {};

\node (L) at (2,2) {\Large $\mathcal NC$};
\node (NS) at (-.5,2) {\Large $\mathcal NS$};

\node[above left] (e) at (1.5,4.5) {\Large \color{quantumdarkgray} $e^{\text{NS}}$};
\draw[quantumdarkgray!80,dashed, thick] (2,5.25) -- (1.16,4);
\filldraw[quantumdarkgray!80] (1.5,4.5) circle (2pt);
\draw[<->,quantumviolet,thick] (1.8,4.36) -- (1.46,3.85);
\node[rotate=-30] (CF) at (2.7,3.45) {\Large \color{quantumviolet}$\NCF(e^{\text{NS}})$};
\node[thick,draw=quantumdarkgray!80,cross out,inner sep=0pt,minimum width=2pt,minimum height=2pt] (eNC) at (1.16,4) {};
\node[thick,draw=quantumdarkgray!80,cross out,inner sep=0pt,minimum width=2pt,minimum height=2pt] (eNC) at (2,5.25) {};

\end{tikzpicture}}
\end{minipage} \hspace{.8cm}
\begin{minipage}[c]{.45\textwidth}
    \centering
    \scalebox{1}{\makeatletter
\tikzoption{canvas is plane}[]{\@setOxy#1}
\def\@setOxy O(#1,#2,#3)x(#4,#5,#6)y(#7,#8,#9)%
  {\def\tikz@plane@origin{\pgfpointxyz{#1}{#2}{#3}}%
   \def\tikz@plane@x{\pgfpointxyz{#4}{#5}{#6}}%
   \def\tikz@plane@y{\pgfpointxyz{#7}{#8}{#9}}%
   \tikz@canvas@is@plane
  }
\makeatother  

\begin{tikzpicture}[scale=1.]

\tikzstyle{vertex}=[circle, draw, inner sep=0pt, minimum size=3pt]
\newcommand{\vertex}{\node[vertex]}


\vertex[yellow!80,fill=yellow!80] (s1) at (2,-2.5,-5.25) {};
\vertex[yellow!80,fill=yellow!80] (s2) at (2,2.5,-5.25) {};
\vertex[yellow!80,fill=yellow!80] (s3) at (5.25,-2.5,-2) {};
\vertex[yellow!80,fill=yellow!80] (s4) at (5.25,2.5,-2) {};
\vertex[yellow!80,fill=yellow!80] (s5) at (-1.25,2.5,-2) {};
\vertex[yellow!80,fill=yellow!80] (s6) at (-1.25,-2.5,-2) {}; %
\vertex[yellow!80,fill=yellow!80] (s7) at (2,2.5,1.25) {}; 
\vertex[yellow!80,fill=yellow!80, text=red] (s8) at (2,-2.5,1.25) {}; %

\draw[draw=none,fill=yellow!10,opacity=0.3] (2,-2.5,-5.25) -- (2,2.5,-5.25) -- (0,0,-4) -- (2,-2.5,-5.25);
\draw[draw=none,fill=yellow!10,opacity=0.3] (2,-2.5,-5.25) -- (2,2.5,-5.25) -- (4,0,-4) -- (2,-2.5,-5.25);

\draw[draw=none,fill=yellow!10,opacity=0.3] (2,2.5,-5.25) -- (-1.25,2.5,-2) -- (0,0,-4) -- (2,2.5,-5.25);
\draw[draw=none,fill=yellow!10,opacity=0.3] (2,2.5,-5.25) -- (5.25,2.5,-2) -- (4,0,-4) -- (2,2.5,-5.25);

\draw[draw=none,fill=yellow!10,opacity=0.3] (0,0,-4) -- (-1.25,-2.5,-2) -- (2,-2.5,-5.25) -- (0,0,-4);
\draw[draw=none,fill=yellow!10,opacity=0.3] (4,0,-4) -- (5.25,-2.5,-2) -- (2,-2.5,-5.25) -- (4,0,-4);

\draw[draw=none,fill=yellow!10,opacity=0.3] (-1.25,-2.5,-2) -- (-1.25,2.5,-2) -- (0,0,-4) -- (-1.25,-2.5,-2);

\draw[draw=none,fill=yellow!10,opacity=0.3] (5.25,-2.5,-2) -- (5.25,2.5,-2) -- (4,0,-4) -- (5.25,-2.5,-2);

\draw[dashed, opacity=.3]  (2,-2.5,-5.25) -- (2,2.5,-5.25) -- (0,0,-4) -- (2,-2.5,-5.25);
\draw[dashed, opacity=.3]  (2,-2.5,-5.25) -- (4,0,-4) -- (2,2.5,-5.25);

\draw[opacity=.3]  (2,2.5,-5.25) -- (-1.25,2.5,-2);
\draw[opacity=.3]  (2,2.5,-5.25) -- (5.25,2.5,-2);
\draw[dashed, opacity=.3] (0,0,-4) -- (-1.25,2.5,-2);

\draw[dashed, opacity=.3]  (0,0,-4) -- (-1.25,-2.5,-2);
\draw[dashed, opacity=.3]  (-1.25,-2.5,-2) -- (2,-2.5,-5.25);

\draw[dashed, opacity=.3]  (4,0,-4) -- (5.25,-2.5,-2) -- (2,-2.5,-5.25);

\draw[dashed, opacity=.3]  (5.25,2.5,-2) -- (4,0,-4);

\begin{scope}[canvas is plane={O(0,0,0)x(1,0,0)y(0,0,-1)},transform shape]
\draw[quantumturquoise!20,fill=quantumturquoise!20] (0,4) -- (2,5.25) -- (4,4) -- (5.25,2) -- (4,0) -- (2,-1.25) -- (0,0) -- (-1.25,2) -- (0,4);

\draw[quantumturquoise!40,fill=quantumturquoise!40]  (0,0) rectangle (4,4);

\vertex[quantumturquoise!100,fill=quantumturquoise!100] (p1) at (0,0) {};
\vertex[quantumturquoise!100,fill=quantumturquoise!100] (p2) at (0,4) {};
\vertex[quantumturquoise!100,fill=quantumturquoise!100] (p3) at (4,0) {};
\vertex[quantumturquoise!100,fill=quantumturquoise!100] (p4) at (4,4) {};

\vertex[quantumturquoise!60,fill=quantumturquoise!60] (p5) at (2,5.25) {};
\vertex[quantumturquoise!60,fill=quantumturquoise!60] (p6) at (5.25,2) {};
\vertex[quantumturquoise!60,fill=quantumturquoise!60] (p7) at (2,-1.25) {};
\vertex[quantumturquoise!60,fill=quantumturquoise!60] (p8) at (-1.25,2) {};

\node (L) at (2,2) {\Large $\mathcal NC$};
\node (NS) at (-.5,2) {\Large $\mathcal NS$};

\node[above left,rotate=-10] (eNS) at (1.5,4.5) {\Large \color{quantumdarkgray} $e^\text{NS}$};
\draw[dashed, quantumgray!80, thick] (2,5.25) -- (1.16,4);
\filldraw[quantumdarkgray!80] (1.5,4.5) circle (2pt);
\draw[<->,quantumviolet,thick] (1.8,4.36) -- (1.46,3.85);
\node[rotate=-30] (NCF) at (2.7,3.25) {\Large \color{quantumviolet}$\NCF(e^\text{NS})$};
\node[thick,draw=quantumdarkgray!80,cross out,inner sep=0pt,minimum width=2.5pt,minimum height=2.5pt] (eNC) at (1.16,4) {};
\node[thick,draw=quantumdarkgray!80,cross out,inner sep=0pt,minimum width=2.5pt,minimum height=2.5pt] (eNC) at (2,5.25) {};
\end{scope}

\draw[dashed, quantumdarkrose, thick] (2,2.5,-5.25) -- (1.5,0,-4.5);
\node[thick,draw=quantumdarkrose,cross out,inner sep=0pt,minimum width=1pt,minimum height=1pt] (eNC) at (1.5,0,-4.5) {};
\node[thick,draw=quantumdarkrose,cross out,inner sep=0pt,minimum width=1pt,minimum height=1pt] (eNC) at (2,2.5,-5.25) {};
\filldraw[quantumdarkrose] (1.735,1,-4.5) circle (2pt);
\node[left] (e) at (1.735,1,-4.5) { \color{quantumdarkrose} $e$};
\draw[<->,quantumblue,thick] (2.02,1,-4.5) -- (1.8,0,-4.36);
\node[right,rotate=-15] (NSF) at (1.8,0.1,-4.5) { \color{quantumblue}$\NSF(e)$};

\draw[draw=none,fill=yellow!20,opacity=0.3] (-1.25,-2.5,-2) -- (2,-2.5,1.25) -- (5.25,-2.5,-2) -- (2,-2.5,-5.25) -- (-1.25,-2.5,-2);

\draw[draw=none,fill=yellow!20,opacity=0.3] (-1.25,-2.5,-2) -- (-1.25,2.5,-2) -- (0,0,0) -- (-1.25,-2.5,-2);

\draw[draw=none,fill=yellow!20,opacity=0.3] (5.25,-2.5,-2) -- (5.25,2.5,-2) -- (4,0,0) -- (5.25,-2.5,-2);

\draw[draw=none,fill=yellow!20,opacity=0.3] (-1.25,2.5,-2) -- (0,0,0) -- (2,2.5,1.25) -- (-1.25,2.5,-2);
\draw[draw=none,fill=yellow!20,opacity=0.3] (-1.25,-2.5,-2) -- (0,0,0) -- (2,-2.5,1.25) -- (-1.25,-2.5,-2);

\draw[draw=none,fill=yellow!20,opacity=0.3] (5.25,2.5,-2) -- (4,0,0) -- (2,2.5,1.25) -- (5.25,2.5,-2);
\draw[draw=none,fill=yellow!20,opacity=0.3] (5.25,-2.5,-2) -- (4,0,0) -- (2,-2.5,1.25) -- (5.25,-2.5,-2);

\draw[draw=none,fill=yellow!20,opacity=0.3] (2,-2.5,1.25) -- (2,2.5,1.25) -- (0,0,0) -- (2,-2.5,1.25);
\draw[draw=none,fill=yellow!20,opacity=0.3] (2,-2.5,1.25) -- (2,2.5,1.25) -- (4,0,0) -- (2,-2.5,1.25);

\draw[draw=none,fill=yellow!20,opacity=0.3] (-1.25,2.5,-2) -- (2,2.5,1.25) -- (5.25,2.5,-2) -- (2,2.5,-5.25) -- (-1.25,2.5,-2);

\draw[opacity=0.3] (-1.25,-2.5,-2) -- (-1.25,2.5,-2) -- (0,0,0) -- (-1.25,-2.5,-2);

\draw[opacity=0.3] (5.25,-2.5,-2) -- (5.25,2.5,-2) -- (4,0,0) -- (5.25,-2.5,-2);

\draw[opacity=0.3] (0,0,0) -- (2,2.5,1.25) -- (-1.25,2.5,-2);
\draw[opacity=0.3] (0,0,0) -- (2,-2.5,1.25) -- (-1.25,-2.5,-2);

\draw[opacity=0.3] (4,0,0) -- (2,2.5,1.25) -- (5.25,2.5,-2);
\draw[opacity=0.3] (4,0,0) -- (2,-2.5,1.25) -- (5.25,-2.5,-2);

\draw[opacity=0.3] (2,-2.5,1.25) -- (2,2.5,1.25);

\node (MS1) at (1.75,-2.5,1.25) {\tiny $M_1$};
\node (MS2) at (1.75,2.5,1.25) {\tiny $M_2$};
\node (PR) at (1.75,0,1.25) {\tiny PR};
\node (D) at (-0.2,0,0) {\tiny D};

\end{tikzpicture}}
\end{minipage}
    \caption{
    [Left] A schematic representation of a 2-dimensional cut of the non-signalling correlation polytope for the (2,2,2) Bell scenario.
    Inscribed in it, in darker colour, is the local (or noncontextual) polytope of correlations. Vertices of the polytopes are highlighted. The quantum boundary (not represented) would correspond to a convex line between the two polytopes. The noncontextual fraction of $e$ $\NCF(e)$ is represented and we have $\CF(e) \defeq 1 - \NCF(e)$.
    [Right]
    A schematic representation of a cut of the signalling correlation polytope for the (2,2,2) Bell scenario with the non-signalling (NS) polytope and the noncontextual (NC) polytope inscribed in it. Vertices of the polytopes are highlighted. Note that contextual vertices of the non-signalling polytope, unlike noncontextual ones, are not vertices of the signalling polytope.
    For a signalling behaviour $e$, its non-signalling fraction (maximal weight on the non-signalling part $e^\text{NS}$ in a convex decomposition) is represented. Likewise, the noncontextual fraction of $e^\text{NS}$ is represented.}
    \label{fig:polytopes}
\end{figure}
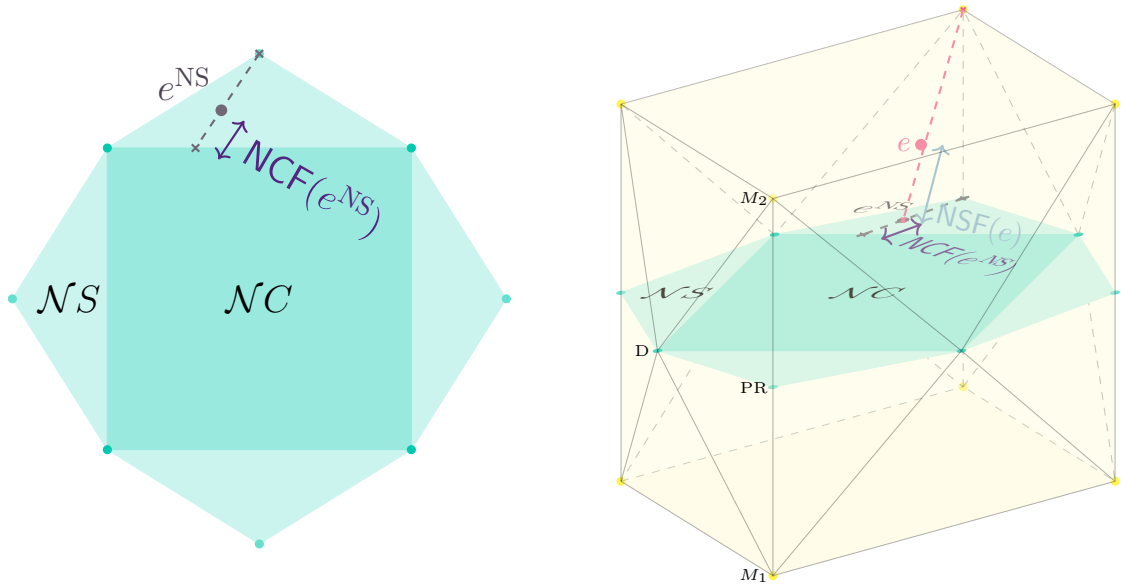

For empirical models outside of the non-signalling polytope, it is possible to define a measure to know \textit{how much the data is signalling} or \textit{how far away from the non-signalling polytope} the empirical model is. This measure has similar properties to the contextual fraction 
but instead of only using global value assignments which are given by the vertices of the noncontextual polytope (see Figure~\ref{fig:polytopes}), we also admit maximally contextual vertices from the non-signalling polytope (e.g.~PR-boxes in the $(2,2,2)$ Bell scenario). To reach those vertices, we allow the global distribution to have negative weight on deterministic noncontextual vertices as long as the marginals for each context are nonnegative. It follows from \cite[Theorems 5.5 and 5.9]{abramsky2011sheaf}.
We call this measure the \textit{non-signalling fraction} $\NSF$.  

\begin{definition}[non-signalling fraction] \label{def:NSF}
Given an empirical model $e$ on a measurement scenario $\XMO$, its non-signalling fraction is given by:
\begin{equation*}
	\NSF(e) \defeq \sup \setdef{\bm 1 \cdot b}{
     \forall C \in \Mc, 0 \leq b|_C \leq e_C}\,.
\end{equation*}
\label{def:nonsign}
\end{definition}
The signalling fraction of $e$ is the given by $\SF(e) \defeq 1 - \NSF(e)$. 
$\NSF$ represents the irreducible weight on signalling vertices one must have in order to explain the data. 
Similar to the contextual fraction, it bears a geometrical interpretation (see Figure~\ref{fig:polytopes})
Asking what fraction of the empirical model $e$ admits a non-signalling explanation is asking for a convex decomposition of the form:
\begin{equation}
	e = \lambda e^\text{NS} + (1-\lambda) e'\,,
\end{equation}
with weight $\lambda$ on the non-signalling part. 
Maximising $\lambda$ yields the non-signalling fraction $\NSF(e)$. 

Idealised models predicted by quantum mechanics, where contexts consist of perfectly commuting observables, should not be signalling;
however, in realistic settings noise can manifest itself to make empirical models somewhat signalling. 
The fraction defined above will capture this effect as a property of the empirical data.
The constraint that for all contexts $C$, $b|_C$ must be nonnegative ensures that, while probabilities might be negative, they sum to give a valid empirical model. Just like the contextual fraction, the signalling fraction of any empirical behaviour can be efficiently computed through a linear program:
\leqnomode
\begin{flalign*}
    \label{prog:LP-SFSupp}
    \tag*{(P-NSF)}
    \hspace{4cm}\left\{
    \begin{aligned}
        & \quad \text{Find } \bm \brm \in \R^n \\
        & \quad \text{maximising } \bm 1.\bm \brm \\
        & \quad \text{subject to:}  \\
        & \hspace{1cm} \begin{aligned}
            & \Mrm \bm \brm \leq \bm \vrm^e \\
            & \Mrm \bm \brm \geq 0 \Mdot
        \end{aligned}
    \end{aligned}
    \right. &&
\end{flalign*}
\reqnomode

Signalling is often associated to disagreeing marginals \cite{guhne2010compatibility,silman2013DeviceIndependent} hence the following quantification:
\begin{definition}[Maximum Incompatibility of Marginals (MIM)] \label{def:MIM}
    The $\MIM$ of an empirical model $e$ in a measurement scenario $\XMO$ is given by the highest incompatibility of marginals:
    \begin{equation}
        \MIM(e) = \max \left \{ \Big\vert \restr{e_C}{C\cap C'} (t) - \restr{e_{C'}}{C\cap C'}(t) \Big\vert \, , \, \forall C \neq C' \in \mathcal{M}, \forall t\in O_{C\cap C'} \right \}\,,
    \end{equation}
    where $O_{C \cap C'}$ is the outcome of a measurement in the intersection of $C$ and $C'$ (see \ref{sec:app_measurementscenario}).
\end{definition}

Both notions of signalling ($\MIM$ and $\SF$) agree on the non-signalling polytope (\ie they are both equal to 0). However there exist points $e$ that are maximally signalling in the sense that they belong to a facet of the signalling polytope thus $\SF(e) = 1$ and for which $\MIM(e) < 1$.
For instance the following empirical model has $\SF = 1$ but $\MIM = 0.2821$:
\begin{equation*}
	\begin{tabular}{cc||cccc}
		 A & B & $00$ & $01$ & $10$ & $11$ \\ \hline
		 $a_1$ & $b_1$ & $0$ & $0$ & $0$ & $1$ \\
		 $a_1$ & $b_2$ & $0.2821$ & $0$ & $0.0674$& $0.6505$ \\
		 $a_2$ & $b_1$ & $0.2821$ & $0.0674$ & $0$ & $0.6505$ \\
		 $a_2$ & $b_2$ & $0.0821$ & $0.4589$ & $0.4589$ & $0$
	\end{tabular}.
\end{equation*}
Given its geometric interpretation, we believe the signalling-fraction is a more natural quantification of signalling.
We prove below that $\MIM$ is always a lower-bound for $\SF$.

\begin{lemma} \label{lemma:MIM_lower_SF}
The signalling fraction of a behaviour $e$ is lower bounded by its $\MIM$.
\end{lemma}

\begin{proof}  
    For any behaviour $e$, by definition of the non-signalling fraction, decompose it as follows:
    \begin{equation}
        e = \NSF(e) e^\text{NS} + \SF(e) e'\,,
    \end{equation}
    with $e^\text{NS}$ a non-signalling model. By convexity of the $\MIM$ measure: 
    \begin{align}
        \MIM(e) 
        &\leq \NSF(e) \MIM(e^\text{NS}) + \SF(e) \MIM(e') \\
        &\leq \SF(e) \MIM(e') \\
        &\leq \SF(e)\,,
    \end{align}
    since $\MIM(e^\text{NS}) = 0$ and $\MIM(e') \leq 1$.
\end{proof}

\subsection{Hidden-variable models}
Hidden variable models (HVMs) provide a broad general approach to considering possible physical explanations underlying observed data,
often with the goal of explaining certain properties in a more intuitive way;
e.g.\ analogous to how probabilistic behaviours of classical statistical mechanics admit a deeper, albeit more complex, deterministic description.
Thus the assumptions that we place on HVMs are usually ways of encoding notions of classicality.

\begin{definition}[Hidden-variable model]
A hidden-variable model on a measurement scenario $\XMO$ consists of the triple $\tuple{\Lambda, p, (h^\lambda)_{\lambda \in \Lambda}}$ where:
\begin{itemize}
    \item $\Lambda$ is the finite space of hidden variables. 
    \item $p$ is a probability distribution on $\Lambda$.
    \item for each $\lambda \in \Lambda$, $h^\lambda$ is empirical model \ie $h^\lambda = (h_C^\lambda)_{C \in \Mc}$ is a family where, for each context $C \in \Mc$, $h_C^\lambda$ is a probability distribution on the joint outcome space $\Oc_C$. 
\end{itemize}
This gives rise to a behaviour $h \defeq \sum_{\lambda \in \Lambda} p(\lambda) h^\lambda$.
\end{definition}

We say that $\tuple{\Lambda, p, (h^\lambda)_{\lambda \in \Lambda}}$ is \textit{parameter-independent} whenever $\forall \lambda$ and for any two contexts $C_1$ and $C_2$, we have that $h^\lambda_{C_1}|_{C_1 \cap C_2} = h^\lambda_{C_2}|_{C_1 \cap C_2}$. This ensures that the behaviour $h$ is non-signalling.

We say that $\tuple{\Lambda, p, (h^\lambda)_{\lambda \in \Lambda}}$ is \textit{outcome deterministic} whenever $\forall \lambda$, $h^\lambda$ gives a deterministic joint outcome for each context (in other words, it is a vertex of the polytope). 

The following proposition is a corollary of \cite[Proposition 3.1 and Theorem 8.1]{abramsky2011sheaf}.
\begin{proposition}
An empirical model $e$ is noncontextual ($\CF(e)=0$) if and only if it is realisable by an outcome deterministic parameter-independent HVM.
\end{proposition}

The set of local behaviour $\mathcal{L}$ is described by a convex combination of hidden variable behaviours which are deterministic and non-signalling. In other words:
\begin{definition}[local polytope]\label{def:local}
The set of local behaviour $\mathcal{L}$ is defined as:
\begin{equation}
    \mathcal{L} \defeq \left\{ h = \sum_{\lambda \in \Lambda} p(\lambda) h^\lambda | \tuple{\Lambda, p, (h^\lambda)_{\lambda \in \Lambda}} \text{ is PI and OD} \right\}\,.
\end{equation}
\end{definition}
This exactly corresponds to the set of behaviours for which the contextual fraction is exactly zero. We start by relaxing determinism. We call $\eta$ the fraction of outcome nondeterminism.
\begin{definition}[$(1-\eta)$ OD HVM] \label{def:ODHVMSupp}
A parameter-independent hidden-variable model is said to be $(1-\eta)$ outcome deterministic if for all hidden variables $\lambda$, there exists a decomposition of $h^\lambda$ of the form
$
    h^\lambda = (1-\eta_\lambda) h^\lambda_\text{OD} + \eta_\lambda h''^{\lambda}
$
with $h^\lambda_\text{OD}$ an outcome deterministic model, potentially signalling, and with necessarily $\eta_\lambda \leq \eta$.
\end{definition}

\noindent This definition ensures that a PR-box \cite{popescu1994quantum} is unsharp since it only admits a $(1-\eta)$ outcome deterministic HVM with $\eta=0.5$ when the hidden variables are parameter-independent. Taking back the measurement scenario for CHSH \eqref{eq:CHSH_measurement_scenario}, a PR-box has a probability table given by Table \ref{tab:PR_box_CHSH}.
\begin{table}[h!t]
	\centering
	\begin{tabular}{cc||cccc}
		 A & B & $00$ & $01$ & $10$ & $11$ \\ \hline
		 $a$ & $b$ & $\sfrac{1}{2}$ & $0$ & $0$ & $\sfrac{1}{2}$ \\
		 $a$ & $b'$ & $\sfrac{1}{2}$ & $0$ & $0$& $\sfrac{1}{2}$ \\
		 $a'$ & $b$ & $\sfrac{1}{2}$ & $0$ & $0$ & $\sfrac{1}{2}$ \\
		 $a'$ & $b'$ & $0$ & $\sfrac{1}{2}$ & $\sfrac{1}{2}$ & $0$
	\end{tabular}
	\caption{PR-box empirical model for the CHSH scenario. This empirical behaviour is parameter-independent and yet it has a contextual fraction of $1$. When all the hidden variables are parameter-independent, the only possible decomposition of the PR-box in a HVM is $e_{\text{PR-box}} = h^\lambda_{\text{PR-box}}$. Thus $\eta = 0.5$ for otherwise there is not parameter-independent HVM describing the PR-box.}
	\label{tab:PR_box_CHSH}
\end{table}

\noindent We now relax parameter-independence:
\begin{definition}[$(1-\sigma)$ PI HVM]\label{def:NSHVMSupp}
A hidden-variable model is said to be $(1-\sigma)$ parameter-independent if for all hidden variables $\lambda$, there exists a decomposition of $h^\lambda$ of the form
$
    h^\lambda = (1 - \sigma_\lambda) h^\lambda_\text{NS} + \sigma_\lambda h'^{\lambda}
$
with $h^\lambda_\text{NS}$ a parameter-independent model and with necessarily $\sigma_\eta \leq \sigma$.
\end{definition}

\noindent If one releases parameter independence then there are many possible decompositions of the PR-box given by Table \ref{tab:PR_box_CHSH}. We give one example of such HVM decomposition with parameter-dependent hidden variable behaviour in Table \ref{tab:HVM_PR_box_1}.

\begin{table}[h!t]%
  \centering
  \subfloat[][$h^\lambda_1$]{
      \begin{tabular}{cc||cccc}
    		 A & B & $00$ & $01$ & $10$ & $11$ \\ \hline
    		 $a$ & $b$ & $1$ & $0$ & $0$ & $0$ \\
    		 $a$ & $b'$ & $1$ & $0$ & $0$& $0$ \\
    		 $a'$ & $b$ & $1$ & $0$ & $0$ & $0$ \\
    		 $a'$ & $b'$ & $0$ & $1$ & $0$ & $0$
	\end{tabular}
  }%
  \qquad
  \subfloat[][$h^\lambda_2$]{
    \begin{tabular}{cc||cccc}
         A & B & $00$ & $01$ & $10$ & $11$ \\ \hline
         $a$ & $b$ & $0$ & $0$ & $0$ & $1$ \\
         $a$ & $b'$ & $0$ & $0$ & $0$& $1$ \\
         $a'$ & $b$ & $0$ & $0$ & $0$ & $1$ \\
         $a'$ & $b'$ & $0$ & $0$ & $1$ & $0$
    \end{tabular}
  }
  \caption{Two hidden variable behaviour such that the HVM given by $h_{\text{PR-box}} = \sfrac{1}{2}h^\lambda_1 + \sfrac{1}{2}h^\lambda_2$ gives a HV explanation for the PR-box. Both hidden variable behaviours are deterministic and parameter-dependent.}%
  \label{tab:HVM_PR_box_1}%
\end{table}

\section{Proof of Theorem 1}

Given two probability distributions $\mu$ and $\nu$ on a countable set $\Omega$, their total variation distance is given by:
\begin{equation}
    V(\mu,\nu) \defeq \frac 12 \sum_{\omega \in \Omega} \vert \mu(\omega) - \nu(\omega) \vert \Mdot
\end{equation}
Fix a (discrete) measurement scenario $\XMO$. Let $e$ and $e'$ be two empirical models on $\XMO$. Define the total variation of a family of distributions by:
\begin{equation}\label{eq:totalvariation}
    V(e,e') \defeq \max_{C \in \Mc} V(e_C,e'_C) \Mdot 
\end{equation}

\begin{theorem}
\label{thm:continuitySupp}
Let $e$ and $e'$ be empirical models on the measurement scenario $\XMO$. If $V(e,e') \leq \varepsilon$ for $\varepsilon > 0$ then $\vert \CF(e) - \CF(e') \vert \leq \vert \Mc \vert \varepsilon $.
\end{theorem}

\begin{proof}
Fix a measurement scenario $\XMO$ and two empirical models $e$ and $e'$ on it.
Let $n \defeq \vert O_X \vert \in \N^*$ the number of global assignments and
let $m \defeq \sum_{C \in \Mc} \vert O_C \vert \in \N^*$ the total number of local assignments ranging over all maximal contexts.

Let $\varepsilon > 0$ and suppose $V(e,e') \leq \varepsilon$. 
Define $\omega = (\omega_C)_{C \in \Mc}$ as $\forall C \in \Mc$, $\omega_C \defeq e_C - e'_C$. The quantity $\omega$ verifies two properties. Because of the normalisation conditions on $e$ and $e'$ we have that:
\begin{equation}
    \forall C\in \Mc, \sum_{o \in O_C} \omega_C(o) = 0 \Mdot
    \label{eq:sum0}
\end{equation}
and because $V(e,e') \leq \varepsilon$ then:
\begin{equation}
    \forall C\in \Mc, \sum_{o \in O_C} \vert \omega_C(o) \vert \leq 2 \varepsilon \Mdot
    \label{eq:varepsboundedness}
\end{equation}
Let $\omega = \omega^+ - \omega^-$ with $\omega^+ = (\omega^+_C)_{C \in \Mc}$ the positive part of $\omega$ and $\omega^- = (\omega^-_C)_{C \in \Mc}$ the negative part of $\omega$. That is, $\forall C \in \Mc$, $\forall o \in O_C$, $\omega^+_C(o) \defeq \max \left\{ 0, \omega(o) \right\}$ and $\omega^-_C(o) \defeq \max \left\{ 0, -\omega(o) \right\}$. Then Eq.~\eqref{eq:sum0} becomes: 
\begin{equation}
    \forall C\in \Mc, \sum_{o \in O_C} \omega^+_C(o) = \sum_{o \in O_C} \omega^-_C(o)\Mdot
    \label{eq:sum0v2}
\end{equation}
And, because $\omega^+$ and $\omega^-$ have disjoint support, Eq.~\eqref{eq:varepsboundedness} becomes: 
\begin{equation}
    \forall C\in \Mc, \sum_{o \in O_C} \omega^+_C(o)  + \sum_{o \in O_C}  \omega^-_C(o)  \leq 2 \varepsilon \Mdot
    \label{eq:varepsboundednessv2}
\end{equation}
Eq.~\eqref{eq:sum0v2} and \eqref{eq:varepsboundednessv2} imply that:
\begin{equation}
    \forall C\in \Mc, \sum_{o \in O_C} \omega^\pm_C(o) \leq \varepsilon \Mdot
    \label{eq:boundedeps}
\end{equation}

We also define $\bm \vrm^{\omega}$ (resp. $\bm \vrm^{\omega+}$ and $\bm \vrm^{\omega-}$) to be the flattened version of the table $\omega$ (resp. $\omega^+$ and $\omega^-$) as defined in the previous section. From, Eq.~\eqref{eq:boundedeps}:
\begin{equation}
    \lVert \bm \vrm^{\omega\pm} \rVert_1 = \bm 1 \cdot \bm \vrm^{\omega\pm} = \sum_{C \in \Mc} \sum_{o \in O_C} \omega^\pm(o) \leq \vert \Mc \vert\varepsilon \Mdot
    \label{eq:boundedepsv2}
\end{equation}

The noncontextual fraction of $e$ $\NCF(e)$ is the optimal value of the linear program:
\leqnomode
\begin{flalign*}
    \label{prog:LP-CFe}
    \tag*{(P-CF$^e$)}
    \hspace{4cm}\left\{
    \begin{aligned}
        & \quad \text{Find } \bm \brm \in \R^n \\
        & \quad \text{maximising } \bm 1.\bm \brm \\
        & \quad \text{subject to:}  \\
        & \hspace{1cm} \begin{aligned}
            & \Mrm \bm \brm \leq \bm \vrm^e \\
            & \bm \brm \geq \bm 0 \Mdot
        \end{aligned}
    \end{aligned}
    \right. &&
\end{flalign*}
\reqnomode
Its dual program reads:
\leqnomode
\begin{flalign*}
    \label{prog:DLP-CFe}
    \tag*{(D-CF$^{e}$)}
    \hspace{4cm}\left\{
    \begin{aligned}
        & \quad \text{Find } \bm \yrm \in \R^m \\
        & \quad \text{minimising } \bm \yrm.\bm \vrm^e \\
        & \quad \text{subject to:}  \\
        &  \hspace{1cm} \begin{aligned}
            & \Mrm^T \bm \yrm \geq \bm 1 \\
            & \bm \yrm \geq \bm 0 \Mdot
        \end{aligned}
    \end{aligned}
    \right. &&
\end{flalign*}
\reqnomode

To compute the noncontextual fraction of $e'$ $\NCF(e')$, it suffices to replace $\bm \vrm^e$ by $\bm \vrm^{e'}$
in the programs above:
\leqnomode
\begin{flalign*}
    \label{prog:LP-CFe'}
    \tag*{(P-CF$^{e'}$)}
    \hspace{4cm}\left\{
    \begin{aligned}
        & \quad \text{Find } \bm \brm \in \R^n \\
        & \quad \text{maximising } \bm 1.\bm \brm \\
        & \quad \text{subject to:}  \\
        & \hspace{1cm} \begin{aligned}
            & \Mrm \bm \brm \leq \bm \vrm^{e' }\\
            & \bm \brm \geq \bm 0 \Mdot
        \end{aligned}
    \end{aligned}
    \right. &&
\end{flalign*}
\begin{flalign*}
    \label{prog:DLP-CFe'}
    \tag*{(D-CF$^{e'}$)}
    \hspace{4cm}\left\{
    \begin{aligned}
        & \quad \text{Find } \bm \yrm \in \R^m \\
        & \quad \text{minimising } \bm \yrm.\bm \vrm^{e'} \\
        & \quad \text{subject to:}  \\
        &  \hspace{1cm} \begin{aligned}
            & \Mrm^T \bm \yrm \geq \bm 1 \\
            & \bm \yrm \geq \bm 0 \Mdot
        \end{aligned}
    \end{aligned}
    \right. &&
\end{flalign*}
\reqnomode

We denote the positive cone in $\R^d$ by $\R^{d+}$ for an integer $d$. The Lagrangian function \cite{barvinok02} $\R^{n+} \times \R^{m+} \longrightarrow \R$ for \refprog{LP-CFe} and \refprog{DLP-CFe} is, for $\bm \brm \in \R^{n+}$ and $\bm \yrm \in \R^{m+}$:
\begin{equation}
    \Lc(\bm \brm,\bm \yrm) = \bm 1 \cdot \bm \brm - \bm \yrm^T(\Mrm \bm \brm - \bm \vrm^e) = \bm \yrm \cdot \bm \vrm^e - \bm \brm^T (\Mrm^T \bm \yrm - \bm 1) \Mdot
\end{equation}
The primal program \refprog{LP-CFe} indeed corresponds to
\begin{equation}
    \Sup{\bm \brm \in \R^{n+}} \Inf{\bm \yrm \in \R^{m+}} \Lc(\bm \brm,\bm \yrm) \,,
\end{equation} 
while the dual program \refprog{DLP-CFe} to 
\begin{equation}
    \inf_{\bm \yrm \in \R^{m+}} \sup_{\bm \brm \in \R^{n+}} \Lc(\bm \brm,\bm \yrm) \Mdot
\end{equation}
We denote $f$ the dual Lagrangian function, that is:
\begin{equation}
    \begin{aligned}
    f : \R^{m+} &\longrightarrow \R \\
    \bm \yrm &\longmapsto \Sup{\bm \brm \in \R^{n+}} \left[ \bm 1 \cdot \bm \brm - \bm \yrm^T (\Mrm \bm \brm - \bm \vrm^e) \right] \,,
    \end{aligned}
    \label{eq:defg}
\end{equation}
which is well-defined under the constraint $\Mrm^T \bm \yrm - \bm 1 \geq 0$. The relation $\geq$ has to be understood as a cone constraint \ie the fact the all coefficients are positive. 

Of course, we can have a similar treatment for $e'$. This yields the dual Lagrangian function:
\begin{equation}
    \begin{aligned}
    f' : \R^{m+} &\longrightarrow \R \\
    \bm \yrm &\longmapsto \Sup{\bm \brm \in \R^{n+}} \left[ \bm 1 \cdot \bm \brm - \bm \yrm^T (\Mrm \bm \brm - \bm \vrm^{e'}) \right]\,.
    \end{aligned}
\end{equation}

Importantly, note that \refprog{LP-CFe} and \refprog{DLP-CFe} (resp. \refprog{LP-CFe'} and \refprog{DLP-CFe'}) are strongly dual (\ie they have the same optimal value) and they both have optimal solutions (since they are finite-dimensional linear programs with a finite value and they have interior points \ie points that strictly satisfy all constraints \cite{barvinok02}).

Let $\bm \yrm^* \in \R^{m+}$ be an optimal solution of \refprog{DLP-CFe} and let $\bm \brm' \in \R^{n+}$ be a feasible point of \refprog{LP-CFe'}. Then:
\begin{align}
    \NCF(e) &= f(\bm \yrm^*) \label{eq:lem1}\\
           & \geq \bm 1 \cdot \bm \brm' - \bm \yrm^{*T} (\Mrm \bm \brm' - \bm \vrm^e) \label{eq:lem2} \\
           & \geq \bm 1 \cdot \bm \brm' + \bm \yrm^* \cdot \bm \vrm^{\omega} \label{eq:lem3}\,,
\end{align}
where Eq.~\eqref{eq:lem1} follows from strong duality; Eq.~\eqref{eq:lem2} follows from the definition of $f$ with the supremum on $\bm \brm \in \R^{n+}$ in Eq.~\eqref{eq:defg}; and Eq.~\eqref{eq:lem3} follows from the fact that $\bm \yrm^* \geq 0$ and that if $\bm \brm'$ is feasible for \refprog{LP-CFe'} then $\Mrm \bm \brm' \leq \bm \vrm^e - \bm \vrm^{\omega}$ and thus $- \Mrm \bm \brm' + \bm \vrm^e \geq \bm \vrm^{\omega}$. Since this holds for every $\bm\brm'$ feasible for \refprog{LP-CFe'} then in particular at an optimal solution $\bm \brm'^{*}$ for which $\bm 1 \cdot \bm \brm'^{*} = \NCF(e')$:
\begin{equation}
    \NCF(e) \geq \NCF(e') + \bm \yrm^* \cdot \bm \vrm^{\omega} \Mdot
    \label{eq:inequalityNCF1}
\end{equation}

Now, symmetrically, let $\bm \yrm'^{*} \in \R^{m+}$ be an optimal solution of \refprog{DLP-CFe'} and let $\bm \brm \in \R^{n+}$ be a feasible point of \refprog{LP-CFe}. Then:
\begin{align}
    \NCF(e') &= f'(\bm \yrm'^*) \\
           & \geq \bm 1 \cdot \bm \brm - \bm \yrm'^{*T} (\Mrm \bm \brm - \bm \vrm^{e'})  \\
           & \geq \bm 1 \cdot \bm \brm - \bm \yrm'^* \cdot \bm \vrm^{\omega} \label{eq:lem3'}\,,
\end{align}
where Eq.~\eqref{eq:lem3'} follows from the facts that $\bm \yrm'^* \geq 0$ and that if $\bm \brm$ is a feasible point for the program \refprog{LP-CFe} then $\Mrm \bm \brm \leq \bm \vrm^{e'} + \bm \vrm^{\omega}$ so that $- \Mrm \bm \brm + \bm \vrm^{e'} \geq - \bm \vrm^{\omega}$. In particular, at an optimal solution $\bm \brm^* \in \R^{n+}$ for \refprog{LP-CFe} for which $\bm 1 \cdot \bm \brm^* = \NCF(e)$:
\begin{equation}
    \NCF(e') \geq \NCF(e) - \bm \yrm'^* \cdot \bm \vrm^{\omega} \Mdot
    \label{eq:inequalityNCF2}
\end{equation}
From Eq.~\eqref{eq:inequalityNCF1} and Eq.~\eqref{eq:inequalityNCF2}, we have that:
\begin{equation}
    \NCF(e) + \bm \yrm^* \cdot \bm \vrm^{\omega-} \geq \NCF(e') \geq \NCF(e) - \bm \yrm'^* \cdot \bm \vrm^{\omega+}\,,
\end{equation}
where we recall that $\bm \vrm^{\omega+}$ is the positive part of $\bm \vrm^{\omega}$ and $\bm \vrm^{\omega-}$ is negative part.

Now having a close inspection at the problem \refprog{DLP-CFe} (resp. \refprog{DLP-CFe'}) gives that $\bm 1 \in \R^{m+}$ is necessarily an upper bound for $\bm \yrm^*$ (resp.  $\bm \yrm'^*$) in the convex cone of positive vectors, \ie $\bm 1 \geq \bm \yrm^*$ (resp. $\bm 1 \geq \bm \yrm'^*$). Indeed, $\bm 1$ trivially satisfies all the constraints and gives a weight of 1 to every component in a context. On the other hand, the optimal solution of \refprog{DLP-CFe} (resp. \refprog{DLP-CFe'}) will give, at worst, a weight of 1 to one specific component in every context (it needs to satisfy the constraints $\Mrm^T \bm \yrm \geq \bm 1$ and $\bm \yrm \geq 0$ while minimising $\bm \yrm \cdot \bm \vrm^{e'}$). 

Then, because of Eq.~\eqref{eq:boundedepsv2}:
\begin{align}
    \bm \yrm^* \cdot \bm \vrm^{\omega}_- & \leq \bm 1 \cdot \bm \vrm^{\omega}_- \leq \vert \Mc \vert \varepsilon \\
    - \bm \yrm'^* \cdot \bm \vrm^{\omega}_+ & \geq - \bm 1 \cdot \bm \vrm^{\omega}_+ \geq - \vert \Mc \vert \varepsilon \, ,
\end{align}
so that:
\begin{equation}
    \NCF(e) +  \vert \Mc \vert \varepsilon  \geq \NCF(e') \geq \NCF(e) -  \vert \Mc \vert \varepsilon \Mdot
\end{equation}
Thus: 
\begin{align}
    & \vert \NCF(e) - \NCF(e') \vert \leq \vert \Mc \vert \varepsilon \\
    & \vert \CF(e) - \CF(e') \vert \leq \vert \Mc \vert \varepsilon\,.
\end{align}
\end{proof}

\section{Proofs of Lemma 1 and Theorem 2}

\begin{lemma}
\label{lemma:correctedSupp}
Let $\tuple{ \{ \lambda \}, \delta_\lambda, (h^\lambda)}$ be a $(1 - \eta_\lambda)$ and $(1 - \sigma_\lambda)$ HVM.
Then,
if $2\eta_{\lambda} + \sigma_{\lambda} < 1$, the contextual fraction of each hidden variable behaviour satisfies $\CF(h^\lambda) \leq \eta_\lambda$.
\end{lemma}
\noindent \textit{Proof.} To prove Lemma~\ref{lemma:correctedSupp} we will decompose $h^\lambda$ according to its deterministic and non-deterministic part optimally:
\begin{equation} \label{eq:optimalODdecomp}
    \exists h^\lambda_\text{OD}, h'^{\lambda}\quad \text{s.t}\quad  h^\lambda = (1- \eta_\lambda^*) h^\lambda_\text{OD} + \eta_\lambda^* h'^{\lambda}\,.
\end{equation}
Decomposing optimally implies that $\eta_\lambda^*$ is irreducible, hence any other decomposition will imply $\eta_\lambda > \eta_\lambda^*$.

The proof is divided into two parts. Firstly, when $h^\lambda_\text{OD}$ is parameter independent, the bound is trivially found. Secondly when $h^\lambda_\text{OD}$ is parameter dependent, it is not possible to respect the condition $\sigma_\lambda + 2\eta_\lambda < 1$.

\subsection*{Case $h^\lambda_{\text{OD}}$ is parameter independent.}

\noindent This case follows from the convexity of the contextual fraction using Eq.~\eqref{eq:optimalODdecomp}. Since $h^\lambda_{\text{OD}}$ is parameter independent and deterministic, it must belong to the noncontextual polytope \cite[Prop. 3.1 and Th. 8.1]{abramsky2011sheaf}, thus its noncontextual fraction is 0. On the other hand the contextual fraction of any model is upper bounded by 1 thus:

\begin{equation}
    \begin{split}
        \CF (h^\lambda) &= \CF\left ((1-\eta_\lambda) h^\lambda_{\text{OD}} + \eta_\lambda h'^\lambda \right )\\
        &\leq (1-\eta_\lambda) \CF(h^\lambda_{\text{OD}}) + \eta_\lambda \CF(h'^\lambda)\\
        &\leq \eta_\lambda \CF(h'^\lambda)\\
        &\leq \eta_\lambda \Mdot
    \end{split}
\end{equation}
Again, this is the tightest bound, since we assumed that Eq.~\eqref{eq:optimalODdecomp} is optimally decomposed.

\subsection*{Case $h^\lambda_{\text{OD}}$ is parameter dependent.}

\noindent In this case we can take $\eta_\lambda < 0.5$ for otherwise $2\eta_\lambda + \sigma_\lambda \geq 1$. Thus if our hidden variable behaviour is of the form \eqref{eq:optimalODdecomp}, we will prove by contradiction that it is not possible to satisfy $2\eta_\lambda + \sigma_\lambda < 1$. To do so, we rely on the fact that the signalling fraction is lower bounded by the maximum incompatibility of marginals (see Lemma~\ref{lemma:MIM_lower_SF}) and we prove the following lemma:

\begin{lemma} \label{lemma:MIM_bound}
    For a decomposition of $h^\lambda$ of the form $(1- \eta_\lambda) h^\lambda_\text{OD} + \eta_\lambda h'^{\lambda}$ (see Eq.~\eqref{eq:optimalODdecomp}) where $h^\lambda_{\text{OD}}$ is parameter dependent and $\eta_\lambda < 0.5$, we have $ \MIM(h^\lambda) \geq 1 - 2\eta_\lambda$.
\end{lemma}

\begin{proof}
    Let us decompose further Eq.~\eqref{eq:optimalODdecomp}:
    \begin{equation}
    \begin{split}
        h^\lambda &= (1 - 2\eta_\lambda) h^\lambda_{\text{OD}} + 2\eta_\lambda (\sfrac{1}{2}\,h^{'\lambda} + \sfrac{1}{2}\, h^\lambda_{\text{OD}}) \\
        &= (1 - 2\eta_\lambda) h^\lambda_{\text{OD}} + 2\eta_\lambda h^{\lambda''} \Mdot
    \end{split}
    \label{eq:optimalODdecomp_extended}
    \end{equation}
    Since $h^\lambda_{\text{OD}}$ is deterministic and parameter dependent, we know that $\MIM(h^\lambda_{\text{OD}}) = 1$, which implies:
    \begin{equation}
        \exists C, C' \in \mathcal{M}, t\in O_{C\cap C'}, \quad \restr{h^\lambda_{\text{OD},C}}{C\cap C'}(t) - \restr{h^\lambda_{\text{OD},{C'}}}{C\cap C'}(t) = 1 \Mdot
        \label{eq:MIM_1_h_OD}
    \end{equation}
    
    Then because $\restr{h^{'\lambda}_C}{C\cap C'}(t) - \restr{h^{'\lambda}_{C'}}{C\cap C'}(t) \geq -1$:
    \begin{equation}
        \sfrac{1}{2}(\restr{h^\lambda_{\text{OD},C}}{C\cap C'}(t) - \restr{h^\lambda_{\text{OD},{C'}}}{C\cap C'}(t)) + \sfrac{1}{2}(\restr{h^{'\lambda}_C}{C\cap C'}(t) - \restr{h^{'\lambda}_{C'}}{C\cap C'}(t)) \geq 0 \Mdot
        \label{eq:mixing_h_OD_h_l_prime_MIM}
    \end{equation}
    That is:
    \begin{equation}
        \restr{h^{\lambda''}_C}{C\cap C'}(t) - \restr{h^{\lambda''}_{C'}}{C\cap C'}(t))  \geq 0 \Mdot
        \label{eq:MIM_hlambda''}
    \end{equation}
    Extending to the general case of Eq.~\eqref{eq:optimalODdecomp_extended} leads to:
    \begin{equation}
    \begin{split}
        \restr{h^\lambda_C}{C\cap C'}(t) - \restr{h^\lambda_{C'}}{C\cap C'}(t) &= \\
        &(1 - 2\eta_\lambda) (\restr{h^\lambda_{\text{OD},C}}{C\cap C'}(t) - \restr{h^\lambda_{\text{OD},C'}}{C\cap C'}(t)) \\
        &+ 2\eta_\lambda (\restr{h^{\lambda''}_C}{C\cap C'}(t) - \restr{h^{\lambda''}_{C'}}{C\cap C'}(t)) \\
        &\geq 1 - 2\eta_\lambda\,,
    \end{split}
    \label{eq:MIM_OD_decomp_extended}
    \end{equation}
    where the last line of Eq.~\eqref{eq:MIM_OD_decomp_extended} is given by equations \eqref{eq:MIM_1_h_OD} and \eqref{eq:MIM_hlambda''}.
    By definition of $\MIM$:
    \begin{equation}
        \MIM(h^\lambda) \geq 1 - 2\eta_\lambda \Mdot
    \end{equation}
\end{proof}
Finally combining Lemma~\ref{lemma:MIM_lower_SF} and Lemma~\ref{lemma:MIM_bound} we obtain:
\begin{equation*}
    \begin{split}
        \sigma_\lambda 
        &\geq \SF(h^\lambda) \\
        &\geq \MIM(h^\lambda) \\
        &\geq 1 - 2\eta_\lambda\\
        &\implies \sigma_\lambda + 2 \eta_\lambda \geq 1\,.
    \end{split}
\end{equation*}
thus we obtain a contradiction with the assumption. \qed

\begin{theorem} \label{thm:CF_boundSupp}
Let $e = \sum_\lambda p(\lambda) h^\lambda $ a behaviour realisable by a $(1-\sigma)$ PI and $(1 - \eta)$ OD HVM $\tuple{\Lambda, p, (h^\lambda)_{\lambda \in \Lambda}}$ such that $\sigma + 2\eta < 1$. Then its contextual fraction is bounded by above: $\CF(e) \leq \eta$.
\end{theorem}

\begin{proof}
    By definition $\sigma_\lambda + 2\eta_\lambda < \sigma + 2\eta$, thus by using the convexity of the contextual fraction and applying Lemma~\ref{lemma:correctedSupp}:
    \begin{equation}
        \begin{split}
            \CF(e) &\leq \sum_\lambda p(\lambda) \CF(h^\lambda) \\
            &\leq \sum_\lambda p(\lambda) \eta_\lambda \\
            &\leq \eta \sum_\lambda p(\lambda) \\
            &\leq \eta\,.
        \end{split}
    \end{equation}
\end{proof}

\section{High signalling and unsharpness}

We rarely expect the case $\sigma_\lambda + 2\eta_\lambda \geq 1$ to be interesting. From an experimental point of view, considering the departure from the quantum set of correlations to be due to noise, it would mean that the experiment is very noisy. Nonetheless we have a look at this case here and prove for the $n$-cycle scenario~\cite{araujo2013all} that we can find a hidden variable model for which $\sigma_\lambda + 2\eta_\lambda \geq 1$ and $\CF(h^\lambda) = 1$. Since the contextual fraction must be upper-bounded by 1, this means that all the contextuality that is witnessed can be attributed to noise.

\begin{lemma} \label{lemma:closest_point}
    Let a hidden variable behaviour $h^\lambda$ be decomposed as:
    \begin{equation}
        h^\lambda = (1-\beta) h^\lambda_{\text{OD}} + \beta h'^\lambda\,,
    \end{equation}
    where we assume that $h'^\lambda$ does not contain $h^\lambda_{\text{OD}}$. If $\beta < 0.5$ we have that $\beta = \eta_\lambda^*$ with $\eta_\lambda^*$ the irreducible weight on the nondeterministic behaviour.
\end{lemma}

\begin{proof}
    The closeness of a point to a behaviour is given by the maximal weight one can put on the point in a convex decomposition that explains the behaviour.
    In other words one can define the closeness $\rho$ between a deterministic point $h^\lambda_{det} \neq h^\lambda_{\text{OD}}$ and the given behaviour $h^\lambda$ with:
    \begin{equation}
        \rho = \sup \left \{r\, |\, r h^\lambda_{det} \leq h^\lambda,\, r\in [0,1], \text{ for any } h^\lambda_{det} \neq h^\lambda_{\text{OD}} \right \} \Mdot
    \end{equation}
     Our assumption leads to $\rho \leq \beta \leq 0.5$. To show this, by contradiction, if $\rho > \beta$ then it means that:
    \begin{equation*}
        \exists h^\lambda_{det} \neq h^\lambda_{\text{OD}}, \rho h^\lambda_{det} \leq h^\lambda, \quad \rho > \beta\,.
    \end{equation*}
    But since $h^\lambda_{det} \neq h^\lambda_{\text{OD}}$ it means that at least for one context they don't have the same support and thus:
    \begin{equation}
        \exists C\in \mathcal{M},\, t \neq t' \in O,\, h^\lambda_{\text{OD},C}(t) = h^\lambda_{det,C} (t') = 1\,.
    \end{equation}
    This implies that $h^\lambda_C(t) \geq 1-\beta$ and $h^\lambda_C(t') \geq \rho$ so summing over this context:
    \begin{equation}
    \begin{split}
        \sum_{s \in O_C} h^\lambda_C(s) &\geq  (1-\beta) + \rho \\
        &> 1\,,
    \end{split}
    \label{eq:sum_prob_greater_than_1}
    \end{equation}
    where the last inequality comes from the assumption that $\rho > \beta$. This contradicts the normalisation of the hidden variable model, thus we have proven that there does not exist $h^\lambda_{det}$ that is closer to $h^\lambda$.
    In other words, writing the nondeterministic maximal weight yields
    \begin{equation}
        h^\lambda = (1-\eta_\lambda^*)h^\lambda_{\text{OD}} + \eta_\lambda^* h'^\lambda
    \end{equation}
    and since $h^\lambda_{\text{OD}}$ is the closest point it corresponds to the best decomposition.
\end{proof}

\begin{lemma} \label{lemma:noisy_bound}
    For the $n$-cycle scenario, it is possible to find a hidden variable model such that $\sigma_\lambda + 2\eta_\lambda = 1$ and $\CF(h^\lambda) = 1$.
\end{lemma}

\begin{proof}
    Let us define a hidden variable model:
    \begin{equation} \label{eq:decomp_upper_bound}
        \alpha \in [0,1], \quad h^\lambda_{ub} = \alpha h^\lambda_{S_1} + (1 - \alpha) h^\lambda_{S_2}\,,
    \end{equation}
    where $h^\lambda_{S_1}$ and $h^\lambda_{S_2}$ are two deterministic parameter dependent hidden variable behaviours and are chosen such that a uniform mixture of both gives a contextual parameter independent behaviour:
    \begin{equation} \label{eq:P_decomp_PR}
        h^\lambda_{\text{CPI}} = \sfrac{1}{2}h^\lambda_{S_1} + \sfrac{1}{2}h^\lambda_{S_2}\,.
    \end{equation}
    In the $n$-cycle scenario it is always possible to find such behaviours~\cite{araujo2013all}, in the CHSH scenario~\cite{clauser1969proposed} this corresponds to the PR-box~\cite{popescu1994quantum}.
    By using Lemma~\ref{lemma:closest_point} on equation~\ref{eq:decomp_upper_bound} we find that $\eta_\lambda = \min(\alpha, 1 - \alpha)$.
    Let us now define another hidden variable behaviour $h'^\lambda_{ub}$:
    \begin{equation} \label{eq:decomp_upper_bound_half}
        \beta \in [0,1], \quad h'^\lambda_{ub} = \beta h^\lambda_{S_1} + (1 - \beta) h^\lambda_{\text{CPI}}\,.
    \end{equation}
    We know that $\MIM(h^\lambda_{S_1}) = 1$ and $\MIM(h^\lambda_{\text{CPI}}) = 0$. Moreover, in this case:
    \begin{equation}
        \begin{split}
            \MIM(h'^\lambda_{ub}) &= \beta \MIM(h^\lambda_{S_1}) + (1 - \beta) \MIM(h^\lambda_{\text{CPI}}) \\
            &= \beta\,.
        \end{split}
    \end{equation}
    Using Lemma~\ref{lemma:MIM_lower_SF} we have $\SF(h'^\lambda_{ub}) \geq \beta$.
    But using the convexity of the signalling fraction we find:
    \begin{equation}
        \begin{split}
            \SF(h'^\lambda_{ub}) &\leq \beta \SF(h^\lambda_{S_1}) + (1 - \beta) \SF(h^\lambda_{\text{CPI}}) \\
            &\leq \beta\,.
        \end{split}
    \end{equation}
    Then $\SF(h'^\lambda_{ub}) = \beta$.
    Converting Eq.~\ref{eq:decomp_upper_bound_half}, by inserting Eq.~\ref{eq:P_decomp_PR}:
    \begin{equation*}
        \begin{split}
            h'^\lambda_{ub} &= \beta h^\lambda_{S_1} + \beta (\sfrac{1}{2} h^\lambda_{S_1} + \sfrac{1}{2} h^\lambda_{S_2}) \\
            &= \frac{1}{2} (1+\beta) h^\lambda_{S_1} + \frac{1}{2}(1-\beta)h^\lambda_{S_2} \\
            &= \alpha h^\lambda_{S_1} + (1-\alpha) h^\lambda_{S_2}, \quad \forall \alpha \in [\sfrac{1}{2}, 1]\,,
        \end{split}
    \end{equation*}
    where $\alpha$ is related to $\beta$ by $\alpha = \frac{1}{2}(1 + \beta)$. Again $\eta_\lambda^* = 1 - \alpha$ due to Lemma~\ref{lemma:closest_point}:
    \begin{equation*}
        \begin{split}
            \eta_\lambda^* &= 1 - \alpha\\
            \eta_\lambda^* &= 1 - \frac{1}{2}(1+\beta)\\
            \eta_\lambda^* &= 1 - \frac{1}{2}(1+\sigma_\lambda^*)\\
            2\eta_\lambda^* &= 1 - \sigma_\lambda^*\\
            \sigma_\lambda^* &= 1 - 2\eta_\lambda^*\,.
        \end{split}
    \end{equation*}
    Hence we have found a hidden variable model $h^\lambda$ such that $\sigma_\lambda^* + 2\eta_\lambda^* = 1$.
    
    We now prove that $\CF(h^\lambda_{ub}) = 1$. This proof is based on the fact that $\CF(h^\lambda_{CPI}) = 1$, thus by decomposing it we obtain (according to Eq.~\ref{eq:P_decomp_PR}):
    \begin{equation} \label{eq:CPI_signalling_decomp}
        \begin{split}
            h^\lambda_{CPI} &= \sfrac{1}{2}h^\lambda_{S_1} + \sfrac{1}{2}h^\lambda_{S_2}\\
        &= \frac{\alpha}{2\alpha}h^\lambda_{S_1} + \frac{1}{2}h^\lambda_{S_2} + \frac{1-\alpha}{2\alpha} h^\lambda_{S_2} - \frac{1-\alpha}{2\alpha} h^\lambda_{S_2} \\
        &= \frac{1}{2\alpha} h^\lambda_{ub} + \left ( \frac{1}{2} - \frac{1-\alpha}{2\alpha} \right ) h^\lambda_{S_2}\,.
        \end{split}
    \end{equation}
    Thus by convexity of the contextual fraction one has :
    \begin{equation}
        \begin{split}
            \CF(h^\lambda_{\text{CPI}}) &\leq \frac{1}{2\alpha} \CF(h^\lambda_{ub}) + \left ( \frac{1}{2} - \frac{1-\alpha}{2\alpha} \right ) \CF(h^\lambda_{S_2})\\
            1 &\leq \frac{1}{2\alpha}  \CF(h^\lambda_{ub}) + \left ( \frac{1}{2} - \frac{1-\alpha}{2\alpha} \right )\\
            &\implies \CF(h^\lambda_{ub}) = 1\,,
        \end{split}
    \end{equation}
    where the decomposition is valid only if $\alpha \geq 0.5$. For the case $\alpha \leq 0.5$, one must use $h^\lambda_{S_1}$ instead of $h^\lambda_{S_2}$ in Eq.~\ref{eq:CPI_signalling_decomp}.
    Finally, combining all of the above one gets a hidden variable model with $\sigma_\lambda^* + 2\eta_\lambda^* = 1$ and $\CF(h^\lambda_{ub}) = 1$.
\end{proof}

\end{appendices}

\end{document}